# Plasmons in graphene: Recent progress and applications

Xiaoguang Luo [a], Teng Qiu [a,*], Weibing Lu [b,*], Zhenhua Ni [a,*]

[a] *Department of Physics and Key Laboratory of MEMS of the Ministry of Education, Southeast University, Nanjing 211189, P. R. China*
[b] *State Key Laboratory of Millimeter waves, School of Information Science and Engineering, Southeast University, Nanjing 210096, P. R. China*

**Abstract**

Owing to its excellent electrical, mechanical, thermal and optical properties, graphene has attracted great interests since it was successfully exfoliated in 2004. Its two dimensional nature and superior properties meet the need of surface plasmons and greatly enrich the field of plasmonics. Recent progress and applications of graphene plasmonics will be reviewed, including the theoretical mechanisms, experimental observations, and meaningful applications. With relatively low loss, high confinement, flexible feature, and good tunability, graphene can be a promising plasmonic material alternative to the noble metals. Optics transformation, plasmonic metamaterials, light harvesting *etc.* are realized in graphene based devices, which are useful for applications in electronics, optics, energy storage, THz technology and so on. Moreover, the fine biocompatibility of graphene makes it a very well candidate for applications in biotechnology and medical science.

---

* Corresponding authors.
*E-mail addresses:* tqiu@seu.edu.cn (T. Qiu), wblu@seu.edu.cn (W. B. Lu), zhni@seu.edu.cn (Z. H. Ni).

**Content**

# 1. Introduction

## 1.1. Electron oscillations and plasmonics

Oscillations are, generally speaking, the simple back and forth swings of an object as induced by the driving factor inside or outside, which can be found in nearly all the materials from the vast universe to the tiny molecules or even electrons. They are so usual and important in our living planet that can be well utilized and change our world. Among them, the electromagnetic (EM) oscillation has been paid great attentions. The origin of EM oscillation can trace back to the middle and later period of $19^{th}$ century, at which time Maxwell predicted theoretically the existence of EM wave from the electron oscillations, and Hertz confirmed it experimentally. Then, it was realized that EM wave is always around us in the forms of visible or invisible light. Without delay, the applications of EM oscillation were widely exploited especially in the communication technology. From the Maxwell equation, one can obtain two solutions, which stand for the radiative and collective EM waves, respectively. However, the latter one, also named as plasmon, does not attract enough attention until its extraordinary properties are discovered in recent years, which has formed a new subject of plasmonics.

The phenomenon related to plasmon was firstly reported by Wood [1] in 1902, with the results of uneven distribution of light in a diffraction grating spectrum. However, he cannot give a plausible explanation for this so-called Wood's anomalies. After about 40 years, Fano [2] theoretically revealed in 1941 that the Wood's anomalies relied on the subsequently excited Sommerfeld's type EM waves with large tangential momentum on a metallic surface, which cannot be described by Rayleigh's approximation [3]. Nevertheless, these surface waves are very strongly damped in the transversal direction. On the other hand, in 1879, Crookes [4] reported firstly the fourth fundamental state of matter with the positive ions and negative electrons or ions coexisting, and he called it as "radiant matter". Then, Langmuir studied the oscillations in ionized gases and named the ionized state of matter as plasma [5]. Subsequently, he and Tonks [6] declared another important result that plasmas can sustain ion and electron oscillations and formed a dilatational wave of the electron density. This wave is equivalent to Fano's which can be quantized as plasmas oscillations, *i.e.*, plasmons, with one resonant frequency of plasmons existing in one bulk material. Based on amount of experimental and theoretical work on the origin and implications of characteristic energy losses experienced by fast electrons in passing through foils, Pines and Bohm suggested some of these energy losses are due to the excitation of plasmon which was a collective behavior [7-10], and found that the resonant frequency of plasmon in bulk plasma is $\omega_0 = \left(ne^2/m\varepsilon_0\right)^{1/2}$, where $n$ and $m$ are the electron density and mass respectively and $\varepsilon_0$ is permittivity of vacuum. From more detailed numerical calculations in 1957, Ritchie [11] found an anomalous energy loss happened both at and below the resonant frequency of plasmon when an electron traversed the thin films, the cause of which was suggested to be depending on the interface of the materials. This suggestion was quickly confirmed experimentally by Powell and Swan [12]. Actually, the resonant frequency of plasmon is determined by the restoring force that exerts on the mobile charges when they are displaced from equilibrium, for example by the nearby passage of an electron [12,13]. Following the previous work, Stern and Ferrell studied the plasma oscillations of the degenerate electron gas related to the material surface and firstly named them as surface plasmons (SPs) in 1960 [14]. Consequently, SPs are the collective oscillations of charges at the surface of plasmonic materials. Owing to the heavy energy loss, plasmon inside the materials evanesces severely, but fortunately, it can propagate quite a long distance along the surface.

With the development of this field, researchers have found that, SPs can be excited or coupled with the different quantized energies, *i.e.*, photons, electrons and phonons [15-19]. Taking the photon as an example, SPs can couple with photons and form the composite particles of surface plasmon polaritons (SPPs). Theoretically, the dispersion relationship between the frequency and wave vector for SPPs propagating along the interface of semi-infinite medium and dielectric can be obtained by surface mode solutions of Maxwell's equations under appropriate boundary conditions [20]. The non-radiation solution is the SPs dispersion $k_{SP} = k_0\sqrt{\varepsilon_m\varepsilon_d/(\varepsilon_m+\varepsilon_d)}$, where $\varepsilon_m$, and $\varepsilon_d$ are the relative permittivities of medium and dielectric, and $k_0$ is the wave vector of light in free space. It is should be noted that SPPs cannot be excited by light in an ideal semi-infinite medium. To excite SPPs, some ways must be exploited to make the wave vectors of SPs and light matched, where structures such as prism, topological defect and periodic corrugation can do it well [21]. From the dispersion relation, $k_{SP}$ can be complex with the positive real part standing for propagation and negative one standing for attenuation which always rely on $\varepsilon_m$. For metallic materials, $\varepsilon_m$ can be derived from the Drude model with the result of $\varepsilon_m = 1 - \frac{\omega_0^2}{\omega^2 + i\tau^{-1}\omega}$, where $\tau$ represents the relaxation time of the electrons in metal [22].

It is already known that SPs enable confinement and control of EM energy at subwavelength scales, and the wave excited by the electron oscillations can propagate along the surface of the plasmonic materials [21]. Coupled with electrons, photons or phonons, SPs have promising applications in engineering and applied sciences [23-24]. Thanks to plasmonics, many bottlenecks are broken such as nanophotonics [25,26], metamaterials [27], photovoltaic devices [28], sensors [29] and so on. These EM waves can be excited in the conventional metal materials such as Au, Ag, Cu, Cr, Al, Mg etc., are regarded as the best plasmonic materials in the past for a long time. However, these noble metals suffer large energy losses (*e.g.*, Ohmic loss and radiative loss), moreover, SPs in metals have bad tunability in a fixed structure or device [30,31]. Such shortcomings limit the further development of plasmonics and it is necessary to find new plasmonic materials.

*1.2. Graphene plasmonics*

A revolution of material is coming since graphene was exfoliated successfully from graphite by Novoselov and Geim in 2004 [32]. The electrons in graphene behave like massless Dirac-Fermions, which results in the extraordinary properties, *e.g.*, carriers (both electrons and holes) with ultra high-mobility and long mean free path, gate-tunable carrier densities, anomalous quantum Hall effects, fine structure constant defined optical transmission, and so on [33]. Owing to the two dimensional (2D) nature of the collective excitations, SPs excited in graphene are confined much more strongly than those in conventional noble metals. Moreover, the low losses and the efficient wave localization up to mid-infrared frequencies also lead it to be a promising alternative in the future applications [34-39]. The most important advantage of graphene would be the tunability of SPs, as the carriers densities in graphene can be easily controlled by electrical gating and doping [40-44]. Consequently, graphene can be applied as terahertz (THz) metamaterial [41,42,45,46] and it can be tuned conveniently even for an encapsulated device; other devices such as flexible plasmonic waveguides [47,48], transformation optical devices [43,49] are also exploited recently by utilizing its advantages of great tunability, low loss, flexible nature and so on. Those achievements manifest much more advantages in the control of EM wave compared to the conventional metal materials.

SPs in graphene can be coupled with photons, electrons or phonons. It will form SPPs with photons [39], and composite "plasmaron" particles with electrons [50]. The former one has already been observed in a standing wave mode by infrared nano-imaging recently [51-53]. These experiment results confirm the existence of SPs in graphene and make versatile graphene based plasmonic device applicable and more meaningful. Graphene can definitely enhance the light absorption [54,55] and light can even be completely absorbed with respect to the incident angle in the nanodisk array graphene structure due to the collective effect of graphene SPs [56]. On the other hand, graphene can help to tune the SPs in conventional metals (such as Au) [45,57,58], which makes it a promising plasmonic materials. Some potential applications related to SPs such as graphene based waveguide polarizers [47,59,60] and chemical/biologic sensors [61-63] need to be further exploited.

This review focuses on the recent progress of graphene plasmonics and its applications. Firstly, the mechanism of graphene SPs is reviewed. We emphasize three kinds of coupling forms: SPs with photons, SPs with electrons and SPs with phonons. Secondly, the SPs in graphene and conventional plasmonic materials are compared. Thirdly, some applications of graphene plasmonics such as transformation optics, THz photonic metamaterial, light harvesting, waveguide polarizer, tunability of SPs in metal nanoparticles and biosensor are discussed. Finally, we will draw a conclusion and give perspectives on the future research and applications of graphene plasmonics.

**2. Surface plasmons in graphene**

*2.1. Basic principle of graphene for surface plasmons*

*2.1.1. Electronic structure of graphene*

Unlike the metal plasmonic materials, the one-atom-thick graphene is so thin, $\approx 0.34$ nm, that the semi-infinite interface model cannot be used to describe the plasmonic properties. Graphene was considered to be unstable and cannot exist due to strong thermal fluctuation of 2D materials at the beginning of 19[th] century [64,65]. However, the rise of graphene has come since it was successfully obtained by mechanical exfoliation of graphite and deposited on a Si wafer capped with 300 nm thickness $SiO_2$ [32,66]. Subsequently, the plasmonics based on graphene becomes one of most exciting research topics.

To investigate the plasmons caused by electron oscillations, we firstly introduce the electronic structure of graphene. Single layer graphene, a gapless semiconductor, is a monolayer of carbon atoms packed in a 2D honeycomb lattice with the lattice constant $a \approx 0.142 \text{nm}$. Three $sp^2$ hybridized orbitals are oriented in the *x-y* plane and have mutual 120° angles which causes the honeycomb formation consisting of six covalent $\sigma$ bonds. The remaining unhybridized $2p_z$ orbital is perpendicular to the $x-y$ plane and forms $\pi$ bonds [67], as the atomic structure shown in Fig. 1a. Since each $2p_z$ orbital has one extra electron, the $\pi$ band is half filled. Nevertheless, the half-filled bands in transition elements have played an important role in the physics of strongly correlated systems because of their strong tight-binding character and the large Coulomb energies [33]. From the tight-binding approach when $\hbar = 1$, the energy bands of a single-layer pure graphene from the $\pi$ electrons can be expressed as [33,68]

$$E_{\pm}(\mathbf{k}) = \pm t\sqrt{3 + f(\mathbf{k})} - t'f(\mathbf{k}), \tag{1}$$

where $\mathbf{k}$ is the wave vector, $t$ and $t'$ are the nearest-neighbor and the next nearest-neighbor hopping energy respectively, and $f(\mathbf{k}) = 2\cos(\sqrt{3}k_y a) + 4\cos\left(\frac{\sqrt{3}}{2}k_y a\right)\cos\left(\frac{3}{2}k_x a\right)$. Based on this equation, it is found the obtained two bands are symmetric around zero energy if $t' = 0$, whle for $t' \neq 0$, the electron-hole symmetry is broken, as shown in Fig. 1b. Each carbon atom contributes one $\pi$ electron, the lower band (*i.e.* $E_-$ in Eq. (1)) is completely filled (called as $\pi$ band) and the upper band (*i.e.* $E_+$ in Eq. (1)) is completely empty (named as $\pi^*$ band for the sake of distinction). The two bands touch each other at the Dirac point at each corner of the graphene Brillouin zone, and the band structure close to the Dirac point is cone-like, where the dispersion can be approximately regarded as linear relationship at small wave vector, as shown in the enlarged view of Fig. 1b.

However, there are still $\sigma$ electrons in the lattice which can hardly be described by the tight-binding model. In order to find more information, other approaches such as first-principles should be used to deal with the energy band of graphene. From local-density approximation, Trickey *et al.* [69] obtained the Kohn-Sham energy bands and densities of states (DOSs) of monolayer graphene (Fermi energy $E_F = 0$), as shown in Figs. 1c and d. The $\pi$, $\pi^*$, $\sigma$ and $\sigma^*$ bands have been shown, from which one can estimate the electron hopping energy between themselves, such as $\pi \rightarrow \pi^*$, $\pi \rightarrow \sigma^*$, $\sigma \rightarrow \pi^*$ and $\sigma \rightarrow \sigma^*$ transitions. It is noted that Van Hove singularities exist at the M point of the Brillouin zone with energy difference of about 5 eV. Generally, energy for $\pi \rightarrow \pi^*$ transition is from zero to several electron volts, which nearly corresponds to all of the EM waves that we are interested in. Nevertheless, there are other transitions, for example, $\pi \rightarrow \sigma^*$, $\sigma \rightarrow \pi^*$ and $\sigma \rightarrow \sigma^*$ at higher resonant energy. Combining with DOSs, which are also cone-like similar to the energy band near the Dirac point, the probability of the hopping behavior can also be estimated. It should be noticed that, other elements impacting the results such as doping, substrate, topography *etc.* are not considered here.

For pristine graphene (Fermi level $E_F$ is equal to the energy at Dirac point), there is only one kind of electron-hole excitation (interband transition) at low electron hopping energy because of the empty $\pi^*$ band (conduction band) and the completely filled $\pi$ band (valence band). While for $n/p$-doped graphene, $E_F$ will be away from the Dirac point, which may cause the other kind of electron-hole excitation: intraband transition. Taking the $n$-doped case as an example, as shown in Fig. 2a, $E_F$ is higher than the Dirac point, where $\pi$ band is completely filled and electrons can also be found in the $\pi^*$ band. The electrons both at the bottom of the conduction band and at the top of the valence band can be excited after absorbing a certain amount of energy and momentum. These excitations then form the electron-hole continuum or single-particle excitation (SPE) region in $(\mathbf{q},\omega)$ space for the wave vector $\mathbf{q}=\mathbf{k}-\mathbf{K}$, where $\mathbf{K}$ is the Dirac point in momentum space. Generally, the spectral weight of the allowed excitations in the SPE spectrum is determined by a spectral function of $S(\mathbf{q},\omega)=-\frac{1}{\pi}\mathrm{Im}\Pi(\mathbf{q},\omega)$, where $\Pi(\mathbf{q},\omega)$ is the polarizability function [70], as shown in Fig. 2b. Consequently, the intraband (region I) and interband (region II) transitions in the $n$-doped graphene have distinct boundary in the SPE region, and there are two other regions where the electron-hole excitation is almost restrained.

Graphene is a platform of many-body interactions, in which the charge carriers can interact with other quasi-particles such as photons, phonons, and electrons themselves. When the Fermi energy coincides with the Dirac point energy, electron-electron and electron-phonon interactions inside quasi-freestanding graphene have been confirmed by high resolution angle resolved photoemission spectroscopy [71]. Because of the impacts from many-body interactions, plasmonics in graphene becomes very complicated but absolutely colorful.

*2.1.2. Dispersion relation of graphene surface plasmons*

The dispersion relation of SPs is very important for graphene plasmonics, and numerous achievements have been made both in theory and experiment [34], such as Semi-classical model [72,73], Random-phase approximation (RPA) [74,75], Tight-binding approximation [76,77], First-principle calculation [78], Dirac equation continuum model [79] and electron energy loss spectroscopy (EELS) experiments [78,80] etc. Among them, the Semi-classical model and RPA are commonly used in theoretical analysis, and EELS is very prevalent for experimental study.

*A. Semi-classical model*

The energy-momentum relationship for electrons in graphene is linear over a wide range of energies rather than quadratic, so that the electrons seem like massless relativistic particles (Dirac fermions). As expected, the low energy conductivity (<3 eV) of graphene $\sigma(\omega,\mu,\Gamma,T)$ consists of two parts: intraband and interband contributions, where $\omega$ is radian frequency, $\mu$ is chemical potential, $\Gamma$ is phenomenological scattering rate which is independent of energy $E$, and $T$ is temperature. In macroscopic volume, the thickness of the ultrathin graphene layers can be regarded as infinitesimally thin. The conductivity of graphene can be derived from Kubo formula [81] or RPA [82]. Ignoring the impact of magnetic field (without Hall conductivity), the Kubo formation is [73,81]

$$\sigma(\omega,\mu,\Gamma,T) = -\frac{ie^2(\omega+i2\Gamma)}{\pi\hbar^2}\left[\frac{1}{(\omega+i2\Gamma)^2}\int_0^\infty E\left(\frac{\partial f(E)}{\partial E}-\frac{\partial f(-E)}{\partial E}\right)dE - \int_0^\infty \frac{f(-E)-f(E)}{(\omega+i2\Gamma)^2-4(E/\hbar)^2}dE\right],$$

(2)

of which $f(E) = \{1+\exp[(E-\mu)/k_B T]\}^{-1}$ is Fermi distribution function, and $2\Gamma = \tau^{-1}$ with $\tau \approx \mu_m \mu / ev_F^2$ the electron relaxation time in graphene where $\mu_m$ is the carrier mobility and $v_F \approx 10^6$ m/s is the Fermi velocity [83]. The first term and second term in Eq. (2) correspond to the intraband electron-phonon scattering process and interband electron transition respectively, *i.e.*, $\sigma = \sigma_{intra} + \sigma_{inter}$ [84]. The former is

$$\sigma_{\text{intra}} = i\frac{e^2 k_B T}{\pi \hbar^2 (\omega + i\tau^{-1})}\left[\frac{\mu}{k_B T} + 2\ln\left(e^{-\mu/k_B T} + 1\right)\right], \tag{3}$$

of which the real part contributes to energy absorption or dissipation due to the intraband electrons. It is noted that $\mu \leq E_F$ and $\mu \approx E_F\left[1 - \frac{\pi^2}{12}\left(\frac{k_B T}{\mu}\right)^2\right]$ when $k_B T/\mu$ is very small. The chemical potential of graphene is also determined by the carrier density which is expressed by

$$n = \frac{2}{\pi \hbar^2 v_F^2}\int_0^\infty E\left[f(E) - f(E + 2\mu)\right]dE. \tag{4}$$

For highly doped or gated graphene ($|\mu| \gg k_B T$), $n \approx \frac{\mu^2}{\pi \hbar^2 v_F^2}$ and the chemical potential here can be expressed as $\mu \approx E_F \approx \sqrt{\pi \hbar^2 v_F^2 n}$. In this case, the intraband terms of the graphene conductivity have the Drude-like form, *i.e.*

$$\sigma_{\text{intra}} = \frac{e^2 \mu}{\pi \hbar^2}\frac{i}{(\omega + i\tau^{-1})}. \tag{5}$$

Similarly, the interband contribution of the conductivity has the complex form too. When $|\mu| \gg k_B T$, it can be approximated as [72,85,86]

$$\sigma_{\text{inter}} = \frac{e^2}{4\hbar}\left[\theta(\hbar\omega - 2|\mu|) + \frac{i}{\pi}\ln\left|\frac{\hbar\omega - 2|\mu|}{\hbar\omega + 2|\mu|}\right|\right], \tag{6}$$

where $\theta(\hbar\omega - 2|\mu|)$ is a step function. From these expressions and the experiment of optical absorption of graphene [87], it is easily found that the intraband contribution dominates in the THz and far-infrared region, while in near-infrared and visible region, the interband process is dominating. Anyway, the conductivity of graphene has the complex form $\sigma = \sigma' + i\sigma''$.

With Dyadic Green's functions and Maxwell equation, the behavior of EM wave in the graphene can be described in detail [72,84,88]. It is found that two kinds of EM surface waves can propagate in graphene, transverse electric (TE, *i.e.* p-polarized) and transverse magnetic (TM, *i.e.* s-polarized) surface modes, which is different from the conventional electron system. The imaginary part of conductivity determines which kind of mode can be supported. For $\sigma'' < 0$ and $\sigma'' > 0$, the TE and TM surface waves are supported respectively. These results are also confirmed theoretically by the valid form of the spectrum of EM modes supported by the 2D electron gas layer [89]. In the collisionless limit ($\tau^{-1}=0$) at zero temperature $k_B T/\mu=0$, TM and TE modes are supported by graphene when $0<\hbar\omega/\mu<1.667$ and $1.667<\hbar\omega/\mu<2$, respectively, as shown in Fig. 3a. From numerical simulations, the energy absorption or dissipation happens if $\hbar\omega/\mu>2$, where the real part of the conductivity $\sigma'>0$ (*i.e.* $\sigma'_{inter}>0$ due to $\sigma'_{intra}=0$). However, with the increase of temperature, seen from Fig. 3b, the bifurcation point where $\sigma''=0$ is redshifted slightly at the fixed chemical potential; the real part $\sigma'$ becomes finite at $\hbar\omega/\mu<2$ and causes a finite damping, where the TE mode suffers more heavily than TM one [89]. Furthermore, for carrier concentrations of $10^{11}\sim10^{14}$ cm$^2$, one can also obtain that TM mode locates in the THz and far-infrared regions and TE mode locates in the far-infrared and near-infrared regions. The TM and TE mode SPs in graphene are summarized in Table 1.

The SP dispersion relation can be obtained from the mode of infinite graphene lying in the plane of the interface between two different mediums characterized by relative permeability and dielectric constant, *i.e.* $(\mu_r,\varepsilon_r)$ and $(\mu'_r,\varepsilon'_r)$ respectively. Generally, the impact of magnetism is not considered, so that $\mu_r=\mu'_r=1$. With Maxwell equation [34,90] or Dyadic Green's function [72], the dispersions of TM and TE surface model in graphene can be obtained, *e.g.*, from Maxwell equation, the dispersion relation for TM modes are [34]

$$\frac{\varepsilon_r}{\sqrt{k_{TM}^2-\frac{\varepsilon_r\omega^2}{c^2}}}+\frac{\varepsilon'_r}{\sqrt{k_{TM}^2-\frac{\varepsilon'_r\omega^2}{c^2}}}+\frac{i\sigma}{\omega\varepsilon_0}=0, \qquad (7)$$

where $\omega/c=k_0$. For isolated graphene ($\varepsilon_r=\varepsilon'_r=1$), the dispersion relation of TM modes can be given by

$$k_{TM}=k_0\sqrt{1-\left(\frac{2}{\sigma\eta_0}\right)^2}, \qquad (8)$$

where $\eta_0=\sqrt{\mu_0/\varepsilon_0}\approx377$ Ω is the intrinsic impedance of free space. Certainly, TE

modes can be dealt with in the similar way, and the result in isolated graphene is expressed as

$$k_{\mathrm{TE}} = k_0 \sqrt{1 - \left(\frac{\sigma \eta_0}{2}\right)^2}. \tag{9}$$

In practice, the thickness of the graphene $\Delta$ should be considered and can be brought in by the boundary conditions during the solution of Maxwell equation. As a result, an effective dielectric constant $\varepsilon = \varepsilon_0 + i\sigma/\omega\Delta$ can be obtained. Considering a highly doped graphene on substrate $(\varepsilon_\mathrm{r} \neq 1, \varepsilon_\mathrm{r}' = 1)$ where TM mode is dominating, and using nonretarded approximation [34] or electrostatic limit [39], one can obtain the analytical expression of dispersion relation of SPs: $k_{\mathrm{SP}} \approx i\varepsilon_0 (\varepsilon_\mathrm{r} + 1)\omega/\sigma$. The intraband contribution is dominating, so that Drude-like form of conductivity can be substituted in direct. Note that $\mu \approx E_\mathrm{F}$ because of $|\mu| \gg k_\mathrm{B} T$, the dispersion relation can be obtained as [39]

$$k_{\mathrm{SP}} \approx \frac{\pi \hbar^2}{e^2 E_\mathrm{F}} \varepsilon_0 (\varepsilon_\mathrm{r} + 1)\omega \left(\omega + \frac{i}{\tau}\right). \tag{10}$$

and therefore, the wavelength of SPs in graphene is expressed by:

$$\lambda_{\mathrm{SP}} \approx \lambda_0 \alpha \frac{4E_\mathrm{F}}{\varepsilon_\mathrm{r} + 1} \frac{1}{\hbar(\omega + i\tau^{-1})} \tag{11}$$

Where $\alpha = \frac{1}{4\pi\varepsilon_0} \frac{e^2}{\hbar c} \approx \frac{1}{137}$ is the fine structure constant. Consequently, the dispersion relation and properties of SPs in graphene can be tuned by adjusting the dielectric constant or the chemical potential.

*B. Random-phase approximation*

As early as in 2006, Hwang *et al.* [74] and Wunsch *et al.* [75] had tried to deal with doped graphene by RPA at zero-temperature under self-consistent-field linear response theory. After assuming the relaxation time to be infinite, the 2D polarizability $\Pi(q,\omega)$ and the dielectric function $\varepsilon(q,\omega)$ of graphene can be obtained theoretically. As the model of each electron is assumed to move in the self-consistent field arising from the external field plus the induced field of all electrons, the SPs dispersion $\omega_{SP}(q)$ is determined from the immediate theoretical consequences of $\varepsilon(q,\omega)=0$. In the long-wavelength limit ($q \to 0$), the dispersion of SPs mode for a single-layer graphene can be expressed as

$$\omega_{SP}(q \to 0) = \omega_0 \sqrt{q}, \tag{12}$$

where $\omega_0 = \left(g_s g_\upsilon e^2 E_F / 2\kappa\right)^{1/2}$, $g_s = 2$ and $g_\upsilon = 2$ are the spin and valley degeneracy in graphene [91], and $\kappa$ is the background lattice dielectric constant of the system. From the expression of $E_F$, one can see $\omega_0 \propto n^{1/4}$ for monolayer graphene which is different from the classical 2D plasmons with $\omega_0 \propto n^{1/2}$ though they both have the same dispersion, $q^{1/2}$ [92]. For finite value of $q$, the parabolic trend of dispersion relation disappears, as shown in Fig. 4a. The SPs in region I are not damped, however, they will decay into electron-hole pairs inside the interband SPE continuum because of Landau damping. Moreover, as $q$ increases, the dispersion relation of graphene converges to boundary line of the two kinds of SPE regions but never enters the intraband SPE continuum. The dispersion relation of graphene SPs can also be studied by EELS experiments. Fig. 4b and c show the results of a high-resolution EELS study of low-energy plasmons in a single layer graphene epitaxially grown on SiC [80]. The experiment data are fitted well by the calculation of RPA. Moreover, doped by external potassium atoms, the plasmons can be tuned. The energies of the SPs become higher when the potassium-induced electron density increases.

Although the results can fit some experiments well, RPA can still not perfectly describe the dispersion relation of SPs because of assumption of infinite relaxation time of electrons and the effect of many-body interactions. Plasmons in graphene are very complicated with the dispersion relation of the SPs being different from the RPA results above. For example, it is found the states near the Dirac point interact strongly with plasmons due to the electron-electron interaction [93]. Significantly, Wang and his cooperator [94] found, if the strength of the spin-orbit interaction $\Delta_{SO} \neq 0$, the energy band opened a gap between the valence and conduction bands and between the intraband and interband SPE regions of the semimetal Dirac system. The SPs mode splits into three modes when it enters the interband SPE region for the gaped case. Furthermore, although the temperature goes against the opening, the SPs can locate in gap rather than the interband SPE region, so as to avoid damping. In fact, this phenomenon happens in all the general gaped cases, not only the spin-orbit interaction induced one [95]. As a result, graphene nanoribbons, gated bilayer graphene, strain [96] or other approaches of opening the energy bandgap would be effective to lower the damping loss of SPs.

The other approaches such as tight-binding approximation, first-principle calculation, Dirac equation continuum model *etc.* can also be used to investigate SPs in graphene, and the results are also confirmed in experiments. Some of the results show that the $\pi$ plasmons [97,98] and especially 2D plasmons at long wavelength [99] are acoustic-like with linear dispersion at low energy, which is thought to be due to many-body interactions and the local-field effect [77,98]. Interestingly, the dispersion of the acoustic plasmons at long wavelength is just as the tangent line from origin of the curves obtained by RPA [99]. It is worth noting that SPs in graphene can also be influenced by substrates, and coupling between SPs in graphene and substrate were also investigated [78, 100].

*2.2. Different plasmons in graphene*

It is known that there are two kinds of outer electrons ($\sigma$ and $\pi$ electrons) in graphene. Naturally, they both can support plasmons. For low energy plasmons (also named as 2D plasmons whose energy <3 eV), it is mainly caused by intraband transitions, nevertheless, there are two other kinds of plasmons at higher energies, of which one is named as $\pi$ plasmon and the other is named as $\pi+\sigma$ plasmon. For pristine graphene, only $\pi$ and $\pi+\sigma$ plasmons exist, while 2D plasmons appear in doped graphene. Fig. 5 shows the intensities of electronic excitations by RPA in doped graphene, in the meanwhile, it also shows the dispersion relations of 2D plasmons and $\pi$ plasmons [101]. Like in metals, the best approach to probe plasmons in graphene is EELS, a highly spatially resolved spectroscopy to detect changes in the electronic structure. By recording the energy loss of transmitted or reflected electrons, the electronic structure or the behavior of electrons can be investigated. Other carbon materials have also been studied by EELS, such as graphite [102,103], fullerene [104-106] and carbon nanotubes. $\pi$ and $\pi+\sigma$ plasmons can be easily found in all of these carbon materials, and the plasmons in single wall carbon nanotube (SWCNT) [103,107,108] is close to those in graphene [36,109]. Actually, with radius tending to $\infty$, much of the interpretation of plasmons behavior in SWCNT can be applied to free-standing monolayer graphene [103]. Both $\pi$ and $\pi+\sigma$ plasmons exhibit bulk and surface modes, while for graphene, bulk plasmon mode nearly vanishes. The out-of-plane and in-plane contributions of the SPs split in energy when $d|\mathbf{q}| \to 0$, where $d$ is the thickness of a thin structure, and the out-of-plane mode is forbidden when the $E$ field of a fast moving particle is perpendicular to the plane of graphene. Nevertheless, the very weak out-of-plane mode still exists in graphene because of its non-ignorable thickness and the experimental conditions [102].

At the optical limit, $|\mathbf{q}| \to 0$, plasmons peaks and transition peaks have been observed by EELS, as shown in Table 2 and Fig. 6. For different experimental conditions, the results will be waved slightly. However, from the shown data, it can still be found that the energy at the position of $\pi$ plasmons peak is always less than 10 eV and far less than that of $\pi+\sigma$ plasmons. Furthermore, some other peaks appear in EELS, they should be the resonant transitions of electrons among $\pi$, $\pi^*$, $\sigma$ and $\sigma^*$ bands. Especially, the maximum energy of $\pi \to \pi^*$ transition is smaller but very close to the energy of $\pi$ plasmons which may impact the $\pi$ plasmons heavily. The variation trend of the plasmons peaks from monolayer graphene to graphite is shown in Fig. 6a [102]. By subtracting the zero loss peak under identical conditions, one can see that the peak energy will increase from 4.7 to 7 eV for $\pi$ plasmons and 14.6 to 26 eV for $\pi+\sigma$ plasmons. In order to obtain more information, graphene is transferred onto perforated carbon grids and then investigated by EELS, as shown in Fig. 6b. After subtracted the influence of background, the peaks of $\pi \rightarrow \pi^*$ transition, $\pi$ plasmons and $\pi+\sigma$ plasmons are shown clearly. A linear dispersion relation is obtained at peak energy ($\pi$ plasmons) for monolayer graphene, which is similar to vertically aligned SWCNT but very different from graphite [109], as shown in Fig. 6c.

Unlike the easily-measured high energy ones, low energy 2D plasmons are more difficult to be confirmed. Actually, the many-body interaction effect is very complicated in graphene. For example, from the monolayer graphene formed on TiC(111) surface, Nagashima *et al.* [110] got two separated modes of SPs below 10 eV by EELS, corresponding to 2D plasmons and $\pi$ plasmons. These SPs can described roughly by the classical plasmons theory with $\omega_{SP} = \sqrt{\beta + t_1 q}$ for $\pi$ plasmons and $\omega_{SP} = \sqrt{2\pi e^2 ndq / m^* \varepsilon}$ for the other one [70,110-112], where $\beta$ and $t_1$ are the energy parameters, $n$ and $m^*$ are density and effective mass of electron, $\varepsilon$ and $d$ are dielectric constant and thickness of the graphene layer. The phenomenon of anomalous kink in the 2D plasmons dispersion relation was also found in the epitaxially grown graphene layer on SiC (0001) and was explained by a strong resonance effect in the formation of electron-hole pairs [35,113,114]. Moreover, the dispersion is rather insensitive to defects [113,114]. Similarly at low energy, SPs in graphene on different substrates such as Ir (111) [115], Pt (111) [116], Ni (111) [117] and Au/Ni (111) [118] were studied by EELS. It seems all the experimental data are influenced, more or less, by substrate, doping, temperature or other factors.

Compared to the $\pi$ and $\pi+\sigma$ plasmons at higher energy (>4.5 eV), the 2D plasmons are in THz and infrared region, which is mostly exploited and show potential applications. We will focus on the 2D plasmons in the later part of the review.

### 2.3. Surface plasmons coupled with photons, electrons and phonons

SPs are the collective oscillation of charges, which are silent in plasmonic materials unless drawn by a definite amount of energy and momentum, such as photons, electrons, phonons and so on. Within these interactions, many abnormal phenomena were found gradually in different plasmonic materials so as to color the promising field of plasmonics. As far as we know, photons [39,119,120], electrons [36] and phonons can couple with SPs by the form of quasi-particles [39,50,121,122] which are very interesting in optoelectronic information technique and condensed matter physics.

*2.3.1. Surface plasmon polaritons*

Although EELS and other spectroscopic studies have revealed the existence of SPs in graphene and the interaction between SPs and low-energy electrons or photons [41,102,123-125], the direct visualization of propagating and localized graphene SPs is high desirable. SPs can be excited by photons and form SPPs, which is convenient to be probed in the plasmonics field. Similar to the traditional metals, SPs in graphene also face the mismatch of energy and momentum with those of light in free space. Thus, prism, topological defects and periodic corrugations [21] were adopted to solve this problem. Recently, Fei *et al.* [51,52] obtained the scanning near-field infrared light nanoscopy of SPPs in gated graphene on $SiO_2$ substrate. To experimentally access high wave vectors plasmons, they illuminated the sharp tip of an atomic force microscope (AFM) with a focused infrared beam ($\lambda_0$ =11.2 μm) to let the wave vectors of light match those of plasmons, as shown in Fig. 7a. At the fixed frequency of incident infrared light, the SPPs excited by the illuminated tip can propagate along the sheet, and they will be reflected, interfered and damped at the graphene edges, defects and at the boundary between different layers of graphene (Figs. 7b-e). Owing to the wave properties, the propagated SPPs will be interfered by the reflected ones, and standing waves can be formed if the SPPs evanesce incompletely. The wavelength of SPPs can be conveniently measured from the standing waves rather than the propagating ones, which is ~200 nm, agree with the theoretical prediction of Eq. (11). Moreover, both amplitude and wavelength of SPPs in graphene can be altered by varying the gate voltage.

At the same time, Chen *et al.* [53] obtained the SPPs standing wave in tapered graphene ribbon on the carbon-terminated surface of 6H-SiC by similar approach too, seen from Fig. 8. The SPPs wavelength was measured as $\lambda_{SP} \approx 260$ nm at the incident light of $\lambda_0 = 9.7$ μm. The properties of SPPs can also be successfully tuned by different wavelengths of incident light, dielectric constants of substrate and gate voltages. However, the intensities of fringes in the middle of ribbon are very weak, which implies that the energy loss cannot be ignored in graphene SPPs, even though it is at the weak damping intraband region. It is a big deal of confirming the existence of SPs directly in graphene by optical imaging. The good tunability also let graphene plasmonics be a good alternative for controlling light of subwavelength devices in the future.

There are other approaches to excite SPs in graphene, *e.g.* a dipole emitter. Fig. 9a shows that SPs can strongly enhance the light-matter interactions between a dipole emitter and doped graphene, where the coupling decay heavily by the distance away from the emitter [39]. SPs in graphene can be excited by designing the geometries too, *e.g.* ribbons, nano-disks, antidots[41,42,126]. Periodical structures such as grating can also do this job which seems more meaningful for applications [127,128]. Fig. 9b shows a plasmonic structure of graphene on an etched silicon diffractive gating, from which a sharp notch on the normal-incidence transmission spectra can be obtained because of the coupling between graphene SPs and incident light [127]. Similarly, SPs in graphene can also be excited in the free-standing graphene sheet with electrostatic inhomogeneous periodical doping [129,130].

It is known that the structure of graphene is not always ideal, inhomogeneities such as corrugations, edges, overlaps, defects, impurities *etc*. may be caused intentionally or unintentionally in practical applications. Instead of propagation, SPs in graphene could be localized and in the form of localized surface plasmon resonances (LSPRs). By annular dark-field imaging, structural and chemical information around point defects in graphene can be detected [131], and these point defects can lead to LSPR. Taking silicon and nitrogen atoms point defects as examples [132], see Fig. 10, an obvious field enhancement around those atoms is obtained by EELS imaging. Absolutely, this experiment is a proof of LSPR at the single atom level. At larger scale, SPs can also be localized, and actually, the SPs excited in graphene ribbons, nanodisks, antidots also contains LSPRs, which will be described in detail in Section 2.4.

*2.3.2. Plasmaron*

In addition to light-matter interactions, graphene is the platform for many-body interactions, which impact SPs severely. The SPs study by EELS in graphene is a good example for the electron-electron interactions. Angle-resolved photoemission spectroscopy (ARPES) is also a powerful tool for the study of many-body interactions [133], which can show the detail information of electronic band structure around the Dirac point. For epitaxial graphene on SiC substrate, many-body interactions are very remarkable and are quite often studied [134-136]. When an electron absorb an amount of energy $\hbar\omega$ in highly doped graphene, the decaying processes include transitions, emitting phonons and plasmons, as shown in Fig. 11a. Bostwick *et al.* [135] suggested that the latter two processes causing the renormalized band structure around K point rather than the perfect cone, as shown in the rightmost panel of Fig. 11a. This assumption was further confirmed by direct experimental proofs. By APRES study of epitaxial graphene on SiC (0001) surface, it was found that the band structure around K point is not linear any more but with the kinked structure because of the electron-plasmon interactions, and the effect of phonons, as shown in Fig. 11b. Moreover, the kinked degree is proportional to the electron concentration. Similar data were obtained in single- and few-layer graphene by Zhou *et al.* [136] from APRES too. They described this kinked band structure as an energy gap and proposed that the origin of this gap was the breaking of sublattice symmetry owing to the graphene-substrate interaction. These descriptions are against by Bostwick's group [137] according to their asymmetric data, scanning tunneling microscopy measurements and other theories. Focus on this debate, Polini *et al.* [93] worked out a theoretical result of strong electron-plasmon interactions around K point by RPA and Dirac equation continuum model, where the state scales were at Fermi energy and a characteristic frequency of plasmons depending on the coupling constant was found, which partially explained the ARPES data.

Fig. 11c shows the schematic of electronic excitations in graphene. The electron-plasmon decaying process occurs only in the region where the dispersion of plasmons does not enter the electron-hole pair excitations region [135]. In subsequent detailed ARPES study of doped graphene, Bostwick *et al.* proposed a new composite particle which is named plasmaron to describe such electron-plasmon interactions [50]. This new quasi-particle appears at greater binding energy owing to the extra energy cost of creating a plasmon and electron-hole pair. By the impact of these interactions, the energy bands around K point split up and the Dirac crossing point is changed into a ring ($E_1$) between two points ($E_0$ and $E_2$), as shown in Fig. 11d. The band, passing through the lower energy point ($E_2$), is thought as plasmaron band, which is dampened by defect scattering. While, the energy shift $\Delta E \approx E_0 - E_2$ of the two bands is almost unaltered by defect scattering.

In theory, the plasmaron dispersion on the effective screening of graphene should depend on the underlying substrate [93,99]. Therefore, the band structure around K point should be different from each other at different substrates. To verify this result, graphene was studied by ARPES of graphene on four different structures which are the ($6\sqrt{3} \times 6\sqrt{3}$)$R30°$ reconstruction, the SiC-graphene interface intercalated with gold, hydrogen and fluorine respectively [138]. The experimental data are shown in Table 3, where $\delta E = \Delta E / |E_0 - E_F|$ indicates the energy separation of the pure charge and plasmaron bands, $\delta k = |k_+ - k_-|/|k_F|$ is the momentum separation, $\alpha_G = e^2 / 4\pi\varepsilon_0 \varepsilon \hbar v_F \approx 2.2/\varepsilon$ is graphene effective coupling constant with the graphene effective dielectric constant $\varepsilon$, and $\varepsilon_r \approx 2\varepsilon - 1$ is the substrate effective dielectric constant. These results suggest that substrates impact the bands of pure charge and plasmaron remarkably. Lower dielectric constant will cause strong coupling which imply larger energy and momentum separation of the two different bands. To find more intrinsic physical meaning, Krstajić *et al.* [139] had studied quasi-free-standing graphene within the Overhauser approach by describing the electron-plasmon interaction as a field theoretical problem. It is found that $\Delta E$, on the order of 50-150 meV, increases with the electron concentration which is in agreement with experiments.

*2.3.3. Surface plasmons coupled with phonons*

In the EELS experiments, acoustic-like quasi-linear dispersions of plasmons are found in the long-wavelength limit which is a strange quantum behavior in graphene. As early in 1959, Kohn had predicted that $\partial \omega(q)/\partial q|_{q=2k_F} = \infty$ of the phonon wave mode in ordinary 2D metals due to the strong electron-phonon interactions, which was called Kohn anomaly [140]. By Raman spectroscopy, Kohn anomalies are also found in graphene with a breakdown of the Born-Oppenheimer approximation [141]. Therefore, the phonons interacting with electron might be responsible to the quasi-linear dispersion. Because of the chirality in graphene, Tse *et al.* found four distinct Kohn anomalies (three for longitudinal optical (LO) phonons at $q = \omega/v$ and $2k_F \pm \omega/v$ by RPA, and one for transverse optical (TO) phonons at $q = \omega/v$), which are different from the metals [142]. Moreover, they proposed theoretically that plasmon-phonon coupled modes cannot arise in suspended graphene because of the separate branches of dispersion relation of plasmon and phonon modes.

Following, researchers achieved the breakthrough from graphene on polar substrates. Within the angle-resolved EELS, strong plasmon-phonon coupling ($\approx 130$ meV) was found in epitaxial graphene on SiC(0001) [38]. As interpreted, surface state charges on the SiC are transferred into empty $\pi^*$ states in the graphene sheet, and surface optical phonon modes in SiC cause the $\pi^*$ and $\pi$ electrons in graphene oscillating. Furthermore, a transition from plasmon-like to phonon-like dispersion is obtained with increasing graphene layers, where the discontinuous dispersions of $\omega^\pm$ modes are exhibited, as shown in Fig. 12a. Both modes are strongly damped when they enter into the SPE regions. Combined with EELS data and numerical calculations, a gap in dispersion relation is found between the two modes, where $\omega^+$ modes converge to the LO phonon (*i.e.* surface-optical phonon) dispersion line and $\omega^-$ modes converge to the TO phonon dispersion line [37], as shown in Fig. 12b. The strong coupling with a gap can also be obtained in theory by considering the nonperturbative Coulomb coupling between electronic excitations and phonons [143]. See the results from Figs. 12c-f, the coupling in single-layer graphene is strong at all densities, however, it is strong only at high densities for bilayer graphene, which agree with the EELS results in reference [38]. Consequently, substrates indeed impact the plasmon-phonon coupling seriously [144].

Differ from reference [142], Jablan *et al.* [122] found plasmon-phonon coupling can happen in suspended graphene from the theoretical prediction by using the self-consistent linear response theory. More interestingly, with LP and TP denoting longitudinal (*i.e.* TM) and transverse (*i.e.* TE) plasmons, only the LP-TO and TP-LO coupled modes exist. The two couplings are very different from each other with the former being much stronger than latter. The strength of the LP-TO coupling increases with doping, while it is just the inverse case for the other. In THz and infrared frequency region, two plasmons and four plasmon-phonon coupled modes are found lately [145].

Anyway, plasmon-phonon coupling should be responsible to the observed quasi-linear dispersion of graphene plasmons. In addition to phonons, photons, and electrons from light, substrates, chemical doping or electrical gating, other factors such as magnetic field may also affect graphene plasmons through the many-body interactions. There are still many ambiguities in basic theory and experiment phenomena need to be investigated in this subject.

*2.4. Surface plasmons in graphene with different geometries*

Due to the flexible 2D nature, SPs in graphene can perform versatile properties by various geometry and topography. Many different properties of SPs in graphene were exploited in different structures such as multi-layer [84,146,147], micro/nano-ribbon [41,148], micro/nano-disk/antidot [126], ring [149], and stacks [42] or even corrugation [150] *etc.*, making plasmons in graphene a very promising field in both semi-classical and quantum frameworks. Generally, with the scale of graphene above few dozens of nanometers (*e.g.*, 20 nm for nano-disk, 10 nm for nano-ribbon [151]), the semi-classical theory (*e.g.*, Maxwell equations) is sufficient enough to describe the performance of plasmons. Otherwise, the finite-size and edge effects will play a non-ignorable role and lead to more interesting results for plasmons compared to that in homogeneous graphene. In a word, in addition to the single-layer sheet, other graphene structures are also potential platforms for plasmons. For the separated single-layer graphene arrays or few-layer graphene, SPs are still dominating owing to the ultrathin thickness although the bulk plasmons cannot be ignored any more.

*2.4.1. Surface plasmons in bilayer graphene*

It is known that bi- and multi-layer graphene are also promising materials for many potential applications, including plasmonics. The approaches for studying plasmons in single-layer graphene can be adapted to describe the plasmons in bi- and multi-layer graphene. However, the properties of graphene with different layers will be distinguished from each other because of the different band structures and many-body interactions. The high energy $\pi$ and $\pi+\sigma$ plasmons can also be found in bi- and multi-layer graphene by EELS experiments [146] and the intensities and frequencies of energy peaks increase with the increase of thickness (also see Fig. 6a).

For 2D plasmons in bilayer graphene, Kubo formalism [84,152] can still be exploited with the considerations of intraband and interband contributions, see Eq. (2). For the model of two infinitesimally thin graphene sheets parallel each other, it is found that transverse EM modes are supported in this bilayer graphene with the dispersion of

$$k_{\text{TEM}} \approx k_0 \sqrt{1 - i\frac{2}{\sigma \eta_0 k_0 d}} \quad , \tag{13}$$

where $d$ denotes the distance between the two sheets and the conductivity can also be rewritten by $\sigma = \sigma' + i\sigma''$.

Actually, except transverse EM modes, other ones of TE (TP) and TM (LP) modes can also be supported by bilayer graphene [153-155]. By theoretical calculation, a perfectly nested band structure of Bernal-type stacked bilayer graphene consists of four bands in which the two upper bands are separated by $\gamma \approx 0.4$ eV, as seen from Fig. 13a. Take the electron sufficiently doped case as an instance, at zero temperature, two different situations can be obtained by the relationship between chemical potential and the separated energy. If $\mu < \gamma/2$, both intraband and interband transitions can occur. However, if $\mu > \gamma/2$, exciting an electron-hole pair with $q=0$ (interband transtions) only happens when the energy $\hbar\omega \geq \gamma$, and the special case with $\hbar\omega = \gamma$ is shown by green arrows in Fig. 13a. The conductivity of the bilayer graphene is very different from that of single-layer graphene. When $\mu < \gamma/2$, the interband transitions with $q=0$ start when $\hbar\omega \geq 2\mu$, which is similar to the single-layer case, and subsequently, strong damping comes. However, more extreme points of the imaginary part are found with respect to energy, and a singular point appears when $\hbar\omega = \gamma$, $e.g.$, the case of $\mu = 0.4\gamma$ is shown in Fig. 13b. For $\mu > \gamma/2$, damping happens if $\hbar\omega \geq \gamma$ owing to the interband transitions with $q=0$, $e.g.$, the case of $\mu = 0.9\gamma$ is shown in Fig. 13c. The imaginary part contains the information of plasmons, $i.e.$ $\sigma'' > 0$ for TM modes and $\sigma'' < 0$ for TE modes.

After neglected the electron tunnelling effect between the graphene sheets, the dispersion of plasmons in bilayer graphene can be studied in detail by RPA. Differed from the single-layer case, two intrinsic modes of LPs ($i.e.$ TM modes) are found in the long-wavelength limit [154-156], of which one is the in-phase optical plasmon with the dispersion of $\omega_+ \propto \sqrt{q}$ and the other is the out-of-phase acoustic plasmon with the dispersion of $\omega_- \propto q$ [154-159]. All the properties of these modes depend on the spatially separated distance ($d$) between the two graphene sheets and the electron concentrations of the sheets themselves ($n_1$ and $n_2$). For finite doping, when $d$ increases to infinite value gradually, optical plasmons can couple with acoustic plasmons, and the frequency of the optical plasmons (acoustic plasmons) decreases (increases) slowly until reach to that of plasmons in single-layer graphene. However, the degenerate mode appears only in the long wavelength region rather than the interband SPE region, as shown in Figs. 14a and b. Owing to the four energy bands of bilayer graphene (see Fig. 13a), the SPE regions consist of two different damping mechanisms, of which one is due to the interband electron-hole pairs excitation in the two upper bands ($SPE_{inter}^2$ in Fig. 14) and the other one is the Landau damping ($SPE_{inter}^1$). $SPE_{inter}^{1,2}$ region indicates the overlap of the two damping regions. The increase of imbalance in electron concentrations of two graphene layers (*i.e.*, $n_2/n_1$ decreases) will extend the $SPE_{inter}^2$ region which cause broader damping region for plasmons, as shown in Figs. 14c and d. Consequently, for low loss, it is better to tune the plasmons into the long wavelength region or decrease the imbalance in electron concentrations. Similar to single-layer graphene, a gap between the intraband and interband SPE regions can be opened at some given substrate, temperature, doping, and so on [160-162]. Low loss is also obtained when plasmons are located in this gap.

SPs in these ultrathin sheets are still the dominating components and they can propagate along the sheets in the double-layer structure. Many other properties or couplings of SPs in bilayer graphene are very similar to the single-layer case. However, it's worth mentioning that SPs in the graphene sheets usually couple with each other by the symmetric and antisymmetric modes [163] where the velocity and damping can easily be controlled by gate voltage [164]. Certainly, multilayer graphene may perform many other properties because of the complicate band structures and many-body interactions.

*2.4.2. Surface plasmons in graphene micro/nano-ribbons*

Graphene can easily be tailored into various geometries for practical applications, in which SPs will be different too. The mostly investigated structures are graphene micro/nano-ribbons (GMRs/GNRs). By reducing the degree of freedom, SPPs in these ribbons can propagate in wanted direction, while LSPR is enhanced due to confinement in other directions. As a result, graphene ribbons can be used as wave guides. Waveguide and edge modes can be found in the THz frequency range when SPs propagate along GMR, as shown in Fig. 15a, which are separated from each other by a gap in wave numbers [165]. Moreover, higher frequency or wider ribbon can increase the number of SPs modes, and the propagation length, rather than wave vector, is strongly sensitive to the relaxation time of charge carriers. The LSPR in GMR can also enhance the optical absorption. To strengthen the resonance and increase the absorption area, GMRs arrays are the best choice. For incident EM wave polarized perpendicular to GMRs, the prominent room-temperature optical absorption peaks can be obtained in THz region, and the resonances can also be tuned by electrical doping, incident angle and the array scales [41], which will be discussed in detail in section 4.4. By suppressing transmission in these ribbons, light passing through GMRs arrays can even be completely absorbed [54,166]. Although being approximately proportional to the coverage of graphene, the absorption in GMRs arrays is still stronger than that in the continuous graphene sheet due to the sufficiently high relaxation time.

As the width of graphene ribbon ($w$) decreases to nanometer sizes, i.e. GNRs, a bandgap will be opened (also happens in multilayer ribbons [167]). The gap can reach about 200 meV when $w \approx 15$ nm, and it becomes larger as the width decreases [168-170]. Owing to the band gap, a gap will appears between the intraband and interband SPE regions, so that low loss SPs is hopeful to be obtained in higher frequency region. As $w$ decreases to below 10 nm, the finite-size effects in graphene plasmons become significant so that the classical local EM theory fail to describe [151], as shown in Figs. 15b-d. When $w$ decreases, the energy of plasmon resonance will be increased and the width will be broadened. From numerical data, the plasmons energies and widths are in good agreement with classical local EM theory above ~10 nm and ~20 nm, respectively. For narrower ribbons, plasmons split into several resonances depend on the edges. The effects from zigzag and armchair edges are also very remarkable for plasmons where zigzag edge causes higher-energy and broader plasmon resonance [151,171]. In the undoped case, only the semimetallic armchair GNRs can support the propagation of SPs. The reason is that SPs in both zigzag and armchair ribbons exhibit Landau damping, however, the damping rate for zigzag GNRs is very large so that the collective modes may decay [172,173].

For doped GNRs, SPs with a discrete set of higher-energy optical modes can be obtained and easily controlled by electrical doping [148,151], as shown in Fig. 16a. Strong electric fields are present at different locations across the GNRs, corresponding to the modes of 2D monopole or multipole, where more nodes appear for SPs with higher energy. In GNRs array, SPs modes in different ribbons can couple with each other with the strength and tunability depending on the structure, Fermi energy, dielectric environment and so on. The results of two GNRs hybridization are shown in Figs. 16b and c as an example, where two configurations are considered: vertically offset and coplanar configurations. There are symmetric and antisymmetric couplings of SPs between the two GNRs, both of which would strongly affect the SPs in GNRs.

### 2.4.3. Surface plasmons in graphene micro/nano-disks, -antidots and -rings

To reduce the dimensionality further, graphene with structures of micro/nano-disks, -antidots, -rings are studied [126,149,174,175]. Like metal nanoparticles, the EM field in graphene disk behaves like a dipole. Calculated by RPA and similar to GNRs, finite-size effects cause substantial plasmons broadening compared to the homogeneous graphene when the disk diameter is below ~20 nm [151,171]. Owing to the zero-dimensional nature, LSPR in disk structure becomes very strong, and causes a strong enhanced electrical field. Therefore, the disk structure of graphene might be promising alternative to the metal materials for LSPR. Combined with a nearby quantum emitter (*e.g.* a quantum dot or a molecule), the nonlinear optical response of graphene nanodisks can be obtained [176,177]. Because of the energy transfer and plasmon-plasmon blockage, the optical response of graphene disks can be easily tuned by doping.

The EM field in graphene antidot can also be regarded as a dipole, while plasmons in graphene ring can be treated as the plasmons hybridization (symmetric and antisymmetric ways) from a graphene disk and a smaller diameter antidot, as shown in Fig. 17a. For similar sizes, the energy of the plasmons hybridization for different graphene structures from low to high are: ring (symmetric coupling), disk, antidot and ring (antisymmetric coupling) respectively. Generally, low hybridization energy is beneficial to the plasmons resonance and also the near-field enhancement. The enhancement factor of EM field by plasmons in graphene ring (symmetric coupling) can reach as large as $10^3$ times in THz region, which is almost 20 times larger than similar structure made by gold [149]. Furthermore, the relatively high relaxation times of the charge carriers in these graphene structures enhance the coupling between plasmons and other quasiparticles, and lead to enhanced absorption and suppressed transmission in THz region.

Large-scale patterns with graphene micro-/nano- structures are necessary for practical applications. The SPs coupling between graphene nanostructures on the same plane is relatively weak, however, it is strong in the stacked structures [126,148]. Certainly, there are many pattern techniques that can produce any desirable structures, *e.g.* the graphene/insulator stacks as shown in Fig. 17b [42]. The coupling of plasmons in the stack structures is similar to that in the GNRs. In addition to electrical field, chemical doping, substrate, and magnetic field can also impact SPs in these graphene structures. For instance, with different magnetic field, THz excitation and tunable optical response can be realized in layered graphene structures [178], graphene ribbons [179] and graphene disks [180], respectively. Consequently, magnetic field is also a non-ignored fact for graphene plasmonics.

## 3. Properties of surface plasmons in graphene

The history of plasmonics for metals has been studied for centuries and different applications have been proposed or realized. The possibility of graphene to be the alternative material to the conventional noble metals (such as Au, Ag, Cu, Al *etc.*) in plasmonics relies on its unique properties, which is the focus of the following section.

## 3.1. Basic parameters of surface plasmon polaritons

For the reason of wide application, we focus on the plasmons related to photons, *i.e.* SPPs. Generally, the existence of SPs depends on a negative real part of dielectric constant, *i.e.* $\varepsilon' < 0$ if $\varepsilon = \varepsilon' + i\varepsilon''$. SPs are well pronounced as resonances when $\varepsilon'' \ll -\varepsilon'$ and the losses are very small [181]. These two conditions are also the criteria for the chosen of good plasmonic materials. SPPs propagate along the surface of the plasmonic materials and attenuate both in the direction parallel and perpendicular to the surface. The wavelength ($\lambda_{SPP}$) and propagation distance ($\delta_{SPP}$) can be derived from the complex dispersion relation by taking the real and imaginary parts of the wave vector ($k_{SPP}$), respectively [18,19,39,182]. Combined with the dispersion relationship of SPs, the expressions of wavelength is

$$\lambda_{SPP} = \frac{2\pi}{\mathrm{Re}[k_{SPP}]}. \tag{14}$$

The propagation distance is defined as the distances over which the intensity of the mode decays to $1/e$ of its initial value. Therefore, according to that of the EM field $\propto e^{ikx}$, the propagation distance will be

$$\delta_{SPP} = \frac{1}{2\,\mathrm{Im}[k_{SPP}]}, \tag{15}$$

and the penetration depth in the medium-dielectric semi-infinite system, which describe the exponential fall off of the fields with distance into the two media, can be expressed by

$$\delta_i = \frac{1}{\mathrm{Im}[k_z]} \tag{16}$$

respectively, where $i$ indicates penetrating into mediums (m) or dielectric (d), and $k_z$ is the complex wave vector perpendicular to the interface, *e.g.*, $k_z = k_0\sqrt{\varepsilon_m^2/(\varepsilon_m + \varepsilon_d)}$ inside the mediums and $k_z = k_0\sqrt{\varepsilon_d^2/(\varepsilon_m + \varepsilon_d)}$ inside the dielectric [18,183].

When the mediums are metals, all the parameters can be expressed analytically, as shown in Table 4, where $\varepsilon_m = 1 - \omega_0^2/(\omega^2 + i\tau^{-1}\omega)$ from Drude model. It is clearly shown that $\delta_{SPP} > \lambda_{SPP}$ for metal plasmonic materials. When $|\varepsilon'_m| \gg |\varepsilon_d|$ and the metal is low loss, the propagation distance can be approximated by $\delta_{SPP} \approx \lambda_0 (\varepsilon'_m)^2 / (2\pi\varepsilon''_m)$, from which one can see that a large (negative) real part and a small imaginary part of dielectric constant of metals are beneficial for SPPs [18]. For graphene, the dielectric constant is hard to be expressed directly. The common method is to deal with the conductivity from the Kubo formula. Fortunately, for the highly doped case, the dispersion has a simple Drude-like form (see Eq. (11)) which can be used to describe the properties of SPPs in graphene. From Table 4, it is found that long relaxation time, high Fermi energy and low dielectric constant of substrate benefit the propagation of SPPs in graphene.

### 3.2. Relatively low loss of surface plasmons in graphene

In the visible and infrared parts of spectrum, $|\varepsilon'|$ and $\varepsilon''$ of noble metals increase with the wavelength of incident light, as shown in Table 5 and other relative dielectric constants calculated from reference [184], which suggests that they are the ideal plasmonic materials in visible and near infrared frequencies, and of which Au and Ag are the best choices. Tassin *et al.* [185-187] have made the comparison of the plasmonic properties between metals and graphene. As shown in Fig. 18a (the frequency $f > 20$ THz), graphene is not as good as 30-nm-thick Au film for plasmonics in the visible and infrared spectrum. However, this situation is changed in THz spectrum. The loss of metal plasmonic materials will increase dramatically with the decrease of frequency owing to the increased imaginary part of dielectric constants. Owing to the surface effect, SPPs in metals are confined at the interface with the penetration depth only about several dozens of nanometers in the visible and infrared spectrum, *e.g.*, 20~50 nm for Au and Ag. Nevertheless, the penetration depth will increase rapidly as the frequency decreases to THz region [188], as shown in Fig. 18b, which means more additional loss.

The one-atom thickness graphene does not have this kind of problem in THz.

Compared to noble metals, smaller imaginary part of dielectric constant and penetration depth in THz make graphene loss less [48]. The relatively long optical relaxation time ~$10^{-13}$s (~$10^{-14}$s in Au for comparison) supports the SPPs with a long lifetime, which leads graphene SPPs to dissipate more slowly [148,189-191]. Moreover, the remarkably low loss of graphene SPPs is expected for sufficiently high doping case. As the frequency decreases, Landau damping for plasmons will also be weakened. Therefore, the effective mode index $\delta_{SPP}/\lambda_{SPP}$ increases when the frequency decreases into THz region. It was obtained that $\delta_{SPP}/\lambda_{SPP}$ of the SPPs along the sheet increases up to about $2\times10^3$ when $f=1\,\mathrm{THz}$ from theoretical calculation for free-standing graphene sheet [192]. From the comparison of plasmonic properties between metals and highly doped graphene, we can find that metals are suitable plasmonic materials for visible and near infrared region while highly doped graphene is suitable for far infrared and THz region.

*3.3. High confinement of surface plasmons in graphene*

Another important parameter of plasmonic materials is the confinement of SPPs which describes the ability of confining light. Generally, the confinement of SPPs is valued by the vertical decay length (*i.e.* penetration depth). Simulation results showed that the propagation length of SPPs in graphene increases with the increase of doping concentration in graphene, while that of the vertical decay length decreases. However, see Fig. 18c, even for the case of very high doping (*e.g.* $\mu = 0.8$ eV), the vertical decay lengths of SPPs in graphene are still a few orders smaller than those in the best metallic plasmonic material: Ag [48]. Consequently, the relatively low loss and high confinement of SPPs in graphene show advantages compared to metals, which can support the propagation of SPPs along a flexible and curved surface, which will also be discussed in section 4.2.

Light is highly compressed with the effective SPP index of $\lambda_0/\lambda_{SPP}$. For a metal plasmonic material, the SPP index $\sqrt{\varepsilon_d \varepsilon_m'/(\varepsilon_d+\varepsilon_m')}$ is only slightly larger than $\sqrt{\varepsilon_d}$ because of $-\varepsilon_m \gg \varepsilon_d$, while $\varepsilon_d$ of common dielectrics are not more than 10. Therefore, the SPP indexs of metals are relatively small. As compared, SPPs in highly doped graphene are highly confined with the SPP index of about $\hbar\omega(\varepsilon_r+1)/4\alpha E_F$, where the frequency, substrate, and doping can be tuned as desired. SPP index of 40-70 has been obtained in graphene on different backgrounds [43,52,53], much higher than that of metals and it will be more helpful for applications in subwavelength optics.

*3.4. The tunability of surface plasmons in graphene*

The tunability of SPs in graphene is the most attractive advantage, which will certainly enrich the plasmonics field. SPs can be controlled by structural design of plasmonic materials, however, they cannot be tuned in metals once the structure is fixed. Fortunately, SPs in graphene can be easily tuned by convenient means of doping. As mentioned before, both conductivity of graphene and dispersion of SPs in graphene are related to Fermi energy (or chemical potential at room temperature), frequency and dielectric environment, and they are also influenced by the energy band structure and DOSs. Taking monolayer graphene as an example, the Fermi energy $E_F \approx \mu \approx \sqrt{\pi\hbar^2 v_F^2 n}$ can be easily tuned by changing the charge concentration, which can be realized by external and internal means, *i.e.*, electrical and chemical doping, respectively. It can be applied to both free-standing graphene [32,193-195] and epitaxial graphene [196-198]. Electrical doping, as shown in Fig. 19a, is very convenient to be controlled by electrical and staticelectric gating (*e.g.* ion-gel top gate [41]). Fig. 19b shows the schematic of two different types of doping in graphene by electrical gating: *n*-doping and *p*-doping [66,199]. The internal way of chemical doping is also very important, and especially for composite graphene structure, *e.g.* graphene on SiC is naturally *n*-doped due to charge transfer from the substrate [120]. With a high electropositivity or electronegativity, atoms or molecules can easily dope graphene due to the charge transfer. As a result, *n*-doping of graphene can be easily realized by metal atoms and *p*-doping can be easily realized by polymer molecules formed by nitrogen, oxygen, or fluorine elements. Epitaxial graphene on SiC or metals is usually *n*- or *p*-doped, however, the doping concentration can be easily tuned by the adsorbent atoms or molecules such as Bi, Sb, Au, $H_2O$, $NH_3$, $NO_2$, or the strong electron acceptor of tetrafluorotetracyanoquinodimethane (F4-TCNQ) [199-203]. Take the F4-TCNQ molecule as an example (Fig. 19c), noncovalent functionalization can be formed between graphene and F4-TCNQ via evaporation in ultrahigh vacuum or wet chemistry [160], and electron will be transferred from graphene to F4-TCNQ. It is known that the working frequency of SPs in graphene is commonly in THz region because of the Landau damping at higher energy. However, under high doping, this working frequency of graphene plasmons can be extended effectively to the mid-infrared range. For examples, in theory, the working frequency of SPs can reach up to 80 THz for $\mu = 0.246$ eV and up to 250 THz for $\mu = 0.8$ eV [48].

SPs in doped small scaled graphene will perform different properties owing to the finite-size effect. For instance, the dispersion of SPs in GNRs exposed to electrical field exhibits distinct and spatial profiles that considerably differs from the uniformly doped graphene sheet [204]. Besides electrical and chemical dopings, SPs in graphene can also be tuned by substrates [205], magnetic field and even the temperatures [206]. In a word, there are a lot of ways of controlling the SPs in graphene to meet the need of plasmonic applications.

Tunability of SPs can also happen in graphene/metal hybrid structures [207,208]. It is known that the loss of graphene SPs is very low in the frequency regions from THz to infrared, and it will increase rapidly in the visible region due to the interband transition. Generally, metals are suitable plasmonic materials in visible and near-infrared frequencies, but with bad tunability [31]. After composited with graphene, more control of SPs will be obtained due to EM coupling between graphene and metals. The remarkable effect is revealed in the composited structures of graphene and metal nanoparticles, where the LSPRs of metal nanoparticles can be tuned dramatically by graphene. Take Au nanoparticle as an example [57,58], the frequency of LSRP peak can be tuned by the distance between graphene and metal nanoparticles. In the hybrid graphene-gold nanorod structure, plasmon resonances at optical frequencies can be controlled and modulated by tuning the interband transitions in graphene through electrical gating [207], as shown in Fig. 20. The plasmon resonance will be redshifted or blueshifted depending on the gate voltage. Without doubt, SPs in other noble metals, such as Ag [209], can also be affected by graphene. With the near filed enhancement of LSPRs, these composite structures can be used for tunable SERS in visible frequency [209,210].

Consequently, for the relatively low loss frequency region, *e.g.* from THz to infrared, SPs in graphene can be tuned conveniently by doping. For the visible frequency region, SPs in graphene have no advantage due to the remarkable loss, however, SPs in noble metals can be tuned by coupling with graphene.

## 4. The applications of graphene plasmonics

The extraordinary properties of SPs in graphene, plus its good flexibility, stability and nice biocompatibility make it a good candidate for varies of applications, including electronics, optics, THz technology, energy storage, biotechnology, medical sciences, and so on. Following, we will introduce some meaningful applications of graphene plasmonics.

### 4.1 Tunable terahertz surface plasmons for amplifier, laser, and antenna

It is known that many matters emit THz EM radiations, such as $H_2O$, $CO_2$, $N_2$, $O_2$, CO molecules and so on. THz radiations can unveil intrinsic physical and chemical characters of the matters, so that it can be used in detections, *e.g.*, security check. The nondestructive testing by using low energy THz radiations (<41.3meV) is also very meaningful for biological detection. Moreover, it is a revolution for telecommunication by exploiting THz technology. Unfortunately, the lack of effective THz sources and detectors limits its applications. Recently, graphene with linear band structure near Dirac point and the possibility of opening a bandgap make it a promising material for THz applications [211,212], in which the plasmons might play an important role.

The THz plasmons in highly doped graphene are mainly related to the intraband transitions of electrons. In such case, the Landau damping is weak and the TM modes are dominating. In order to excite THz plasmons, the incident light should match the dispersions of plasmons. In theory, an elementary dipole or quantum emitter can excite THz SPs on graphene surfaces [213-215]. SPs can dominate the response along the suspended graphene sheet and exhibits a strong tunable excitation peak in the THz region. For supported graphene, the interaction is dependent on the dielectric support layer, *e.g.*, the excitation peak will decrease and redshift with poor field confinement on $SiO_2$, and the field confinement will be enhanced for Si. The evanescent wave of incident light can also excite SPs. For instance, with a high-index coupling prism, SPs in doped monolayer graphene can be excited with the frequencies up to about 10 THz, and higher frequencies for few-layer graphene [216]. Due to the enhancement effect, THz SPs in graphene can be used in many fields, such as THz laser [217,218], THz plasmonic antenna [219,220], THz metamaterials [41] and so on.

The conductivity of graphene is complex with the imaginary part implies which mode is supported, while the real part indicates emission or absorption depending on the negative or positive situation, respectively. As calculated, the real part can be negative in THz and infrared frequencies by heavily doping, increasing optical radiation, choosing appropriate substrate and so on [46,218,221]. In addition, the carrier relaxation process in graphene is ultrafast. Picosecond time-scale is obtained in epitaxial graphene layers grown on SiC wafers by using optical-pump THz-probe spectroscopy [189,222]. As a result, it is possible to make ultrafast THz laser based on graphene. In order to make a laser, the interband population inversion need be achieved first. In doped or undoped graphene, femtosecond population inversion [223] can be easily realized by carrier injection [224], plasmons [46], or optical pumping [225,226]. Taking plasmons as an example, see Fig. 21a, graphene can absorb plasmons and then results in interband population inversion in THz region. During the population inversion, the coupling of the plasmons to interband electron-hole transitions in graphene can lead to plasmon amplification through stimulated emission [46]. Similar to optical gain, plasmon gain is also an important parameter. Owing to the slow group velocity and strong confinement of graphene plasmons, graphene oscillator exhibits as a THz plasmonic amplifier with the plasmon gain values being much larger than the typical gain values in semiconductor interband lasers, and most importantly, it can work in THz frequencies. Besides the interband gain, the intraband absorption will lessen the gain. Nevertheless, the optimized working environment can still be found with high gain through adjusting the parameters, *e.g.* doping, substrate, and so on. After defined as the difference between interband gain and intraband loss, the net plasmon gain will increase at a fixed frequency and the frequency range of the plasmon gain will be broadened when the carrier density increases [46]. Similar results can also be obtained in optically pumped graphene structures, as shown in Fig. 21a, where the absorbance can exhibit negative value and the substrate impact them a lot [218]. In addition, the number of graphene layers can also impact the gain. The plasmon amplification weakens with the increase of number of graphene layers.

Nevertheless, a strong dephasing of the plasmon mode is caused by the large plasmon gain, and will prevent THz lasing. In addition, the strong coupling between the plasmons and EM radiation will also hinder the lasing from nonequilibrium plasmons. Facing these situations, Popov *et al.* [217] predicted a planar array of graphene resonant micro/nano-cavities for THz laser, as shown in Fig. 21b. Thanks to the strong confinement of the plasmons and the superradiant nature of EM emission from the cavities, the amplification of THz waves at the plasmon resonance frequency will be several orders of magnitude stronger than that away from the resonances. Furthermore, the plasmons coherence restore in the graphene micro/nano-cavities and couple strongly to the THz radiation at the balance between the plasmon gain and plasmon radiative damping.

Due to the tunability and the strong confinement of the SPs in graphene, other tunable plasmonic devices can also be realized. Combined with the SPs properties of strong coupling with THz EM wave, THz plasmonic antennas can be made. For instance, Tamagnone *et al.* [219,220] proposed a dipole-like plasmonic resonant antenna in a graphene/$Al_2O_3$/graphene stack structure whose properties can be tuned via electrical field, as depicted in Fig. 21c. It is noted that the silicon lens in structure will lead to higher directivity but has negligible impact on the antenna input impedance.

*4.2 Plasmonic waveguide, one-atom thick Luneburg lens, modulator, and polarizer*

SPs in graphene can couple with light (photons) and form the quasiparticles of SPPs, which can propagate along the graphene surface. Devices such as waveguides and lens can be realized in graphene according to the propagation properties of SPPs. Due to the 2D nature, graphene can support both TM and TE modes. The supported modes in the suspended graphene should satisfy the spectrum forms for propagating along and localized near the 2D electron gas layer [89], those are

$$1 + \frac{2\pi i \sigma(\omega)\sqrt{q^2 - \omega^2/c^2}}{\omega} = 0 \quad (15)$$

and

$$1 - \frac{2\pi i \omega \sigma(\omega)}{c^2 \sqrt{q^2 - \omega^2/c^2}} = 0 \quad (16)$$

for the TM and TE waves respectively. From calculation, TE modes locate in the spectrum region from near-infrared frequencies which is dependent on the carrier concentration, while TM modes is supported in the region from infrared to THz frequencies. In practical applications, different factors would impact the dispersion of SPs, see Eq. (8) for the highly doped case and section 2.4 for different structures of graphene. Therefore, to design graphene based waveguides, different approaches might work, with two of them are from substrate and graphene structures.

The one-atom thick monolayer graphene sheet can be used as waveguide because of the strong confinement of SPs. Different chemical potentials will lead to different conductivities and determine whether SPs can propagate. Hence, the wave path of SPs can be controlled by electrical gating with different structured dielectric spacer which holds graphene [43], as shown in Figs. 22a and b. Due to the inhomogeneous permittivity distribution of the spacer near the graphene sheet, a nonuniform static electric field will lead to inhomogeneous chemical potential in graphene sheet and the SPPs wave path can be controlled. It is noted that the edge field of wave path is caused by a gap in wave numbers of the waveguide and edge modes. In a similar way, the structured graphene will also lead to inhomogeneous conductivity. Consequently, waveguides can be designed by tailoring graphene into micro/nano-ribbons, or other structures [47,148,165], also see Figs. 15 and 16. Waveguides designed by these two approaches can both be controlled by external electric field or even magnetic field. In addition to the selection of the supported modes, it is reported that the plasmon velocity can be changed over two orders of magnitude by magnetic field and electric field [227].

Based on the nature of transformation optics, graphene can be used to design other interesting devices. For instance, it can be made into a one-atom thick Luneburg lens, a 2D disk lens with no aberrations for which the locus of focal points resides on a circle. The approach is to achieve several graphene rings with specific conductivity values, where the needed conductivity values can be obtained by external gate voltage or chemical doping [43, 49, 90], as shown Fig. 22c. In addition to waveguide and Luneburg lens, different monolayer graphene plasmonic devices could also be designed by similar manners.

Owing to the high confinement, SPPs can propagate along graphene with low loss even in the curved-plane graphene sheet, which is more difficult to be realized in metal materials [48]. As a demonstration, $180^o$ bending, S-shaped, and spiral waveguides were designed on graphene, with very small energy leakage. Moreover, Y-shaped waveguides and Luneburg lens can be realized on the flexible and curved graphene surface, as shown in Figs 22d, e and f. For multilayer graphene, the coupling of the plasmons between different layers cannot be neglected when interlayer distance is small enough [147,163]. Fig. 22g shows the results of a graphene waveguide splitter by utilizing such coupling. SPPs in different sheet cannot couple with each other when the conductivities of the graphene sheets are different. The splitter can then works as a SPPs switch by gating one of the output graphene sheet. No back scattering is found at the junction of the splitter, and the insertion loss can be avoided.

A pristine graphene monolayer has a constant absorption of about 2.3% across the infrared and visible range due to its unique band structure. However, the absorption will be changed according to the Fermi level in the doped graphene, *e.g.*, it can be controlled by electrical gating. A graphene based optical modulator was realized by covering a graphene sheet on a Si waveguide and applying different gate voltage [228-230]. Gated graphene can also be used as a tunable plasmonic modulator [231]. By RPA, Andersen [232] found that the TM mode plasmons can be absorbed by graphene, where the absorption can be controlled by the applied gate voltage (*i.e.* electroabsorption), as shown in Figs. 23a, b and c. From the simulation results, it is found that higher loss of the plasmons happens at lower carrier concentration due to the enhancement of interband absorption. With this modulator, more than 60 dB of on/off ratio could be obtained for $\lambda_0 = 10\,\mu m$ incoming wave in theory.

Moreover, by tuning the conductivity with electrical gate, graphene can be an optical polarizer [233]. By embedding graphene ribbons in polymer with low refractive index and low loss, they perform as a plasmonic waveguide and polarizer simultaneously which supports the $\lambda_0 = 1.31\,\mu m$ TM polarized mode and with the averaged extinction ratio of 19 dB [47]. Although it is still questionable whether this is a real "plasmonic" device as SPs with such a high frequency may not survive in a commonly doped graphene (see section 3,4). Similar to the design of modulators, the polarizer can be fabricated by the structure of graphene on the top of waveguide directly [59,60]. Fig. 23d shows an in-line fiber-to-graphene polarizer on a side-polished optical fiber, which is attributed to the differential attenuation of the TM and TE polarization modes. From the results, the polarizer can work in a broadband of the spectrum, and a maximum extinction ratio of 27 dB for $\lambda_0 = 1.55\,\mu m$ TM polarized mode was obtained [60].

With graphene plasmonic devices of lasers, waveguides, antennas, modulators, polarizers *etc.*, the plasmonic circuit could also come true. Owing to the strong light-matter interaction, broadband operation, high-speed operation, compatibility with standard metal-oxide-semiconductor processing [228] and high tunability, graphene will dramatically promote the development of nanophotonics, nanoelectronics, plasmonics, and so on.

*4.3. Graphene plasmonic metamaterials*

Metamaterials have been rising for several decades and exhibit peculiar and unnatural properties, where plasmons play important roles. Graphene can also be used to build plasmonic metamaterials, see Fig. 24. Differ from the conventional 2D electron gas system, a prominent room-temperature THz absorption peaks is observed in graphene micro-ribbons array [41], as shown in Fig. 24e. With an ion-gel top gate, this metamaterial can be controlled electrically. In addition, the width of the ribbon will affect the results dramatically. At some appropriate values of width smaller than the wavelength of the input light, the local EM resonances (*i.e.* LSPRs) of GMRs will be enhanced and even result in complete absorption [166]. Stacked structures can enhance the resonance further, and the transmittance can be improved by few-layer stacked graphene micro-disks [42]. There are many other shapes and antishapes for graphene metamaterials, such as antidots, crosses, anticrosses (shown in Fig. 24f) and so on [234]. Anyway, with different structures, graphene can be regarded as a versatile metamaterial for transmission, absorption, modulator, polarizer, or even for the mysterious cloaking [235].

Another important contribution of graphene is that it can tune the properties of conventional metamaterials. The general way is to cover graphene on other metamaterials. Taking Fano rings as examples, see Fig. 24a, it was found that the coverage of graphene can lead to ~250% enhancement of transmission of the metamaterials at the resonant frequency, which was interpreted by the renormalization of the plasmonic modes and the frequency shift of the trapped-mode transmission resonance in the presence of graphene [45]. Figs. 24b and c show the results of numerical simulations of these Fano rings [45,236], where two resonances or dips are found in the transmission, absorption, and reflectance spectra. For transmission spectra, two resonances are corresponding to the dipole resonances which confine the EM wave in I shape (right panel of Fig. 24b) and U shape (not shown) of the ring, respectively, and the dip between them is corresponding to the trapped-mode (left panel of Fig. 24b). It is shown that the transmittance of the composited structure can be higher than that of the metamaterials, and that of the graphene [237]. Moreover the absorbance and reflectance of the metamaterials can also be improved at the specific frequency due to the presence of graphene. In addition to infrared frequencies, this change can happen in other part of the spectrum. For example, a layer of hexagonal metallic meta-atoms covered by monolayer graphene sheet can do this work in THz region [238], see Fig. 24d. Combined with the tunability of graphene, the properties of metamaterials can be further controlled by graphene.

*4.4. Plasmonic light harvesting*

The relatively poor light absorbance of graphene has limited its applications in photodetector and solar cell (photovoltage). This problem can be solved by plasmons, where the LSPR can efficiently enhance the optical absorption at resonant frequency. Many discontinuous graphene structures can be designed and some of them exhibit high efficiency for light absorption due to LSPR. The micro-/nano-ribbons array metamaterials are good examples, where THz absorption peaks are found for the input light polarized perpendicular to the ribbon length [41]. The size and the carrier concentration impact dramatically the absorbance because of the changes of plasmon resonances. Generally, higher frequency absorption peak appears in smaller size structure. From theoretical calculations, it is found that some resonances appear in the absorption spectrum, where the rich ones locate in the frequency ranges from infrared to microwave [54,166]. By adjusting the ribbons size, the substrate and the chemical potential, the transmission and reflection can be completely suppressed while the absorbance can even reach 100%. However, the ribbon still exhibits the intrinsic 1D nature, in which the absorption efficiency is not good enough for the input light polarized in other directions.

For the sake of absorbing input light with different polarizations, the zero-dimensional structures of antidot and disk have been studied [56,174,239]. Similar to metal nanoparticles, LSPRs are stronger in these graphene structures. For the discontinuous graphene sheet with periodic antidot array, see Fig. 25a and b,

LSPR enhances absorption and suppresses transmission more efficiently because of the higher relaxation times of charge carriers. These rich absorption peaks still appears in THz frequencies, however, they will blueshift with the decrease of the antidot size [174]. The converse structure of micro/nano-disk can also have similar properties. Fig. 25c shows the complete optical absorption of periodic graphene nano-disks array, which is valid for input light polarized in all directions [56]. With the effect of LSPR, many absorption peaks will appear in a fixed structure. The maximum absorbance can reach 100% which is dependant on the incidence angle of the input light. As calculated, smaller size of the disk is beneficial for higher maximum absorbance and the range of incidence angles to fulfill the maximum absorbance is allowed to be larger.

There are many other graphene structures which exhibit efficient absorption. However, more attentions are paid on continuous graphene sheet due to the fine electrical properties and convenient fabrication. The LSPR in graphene can be also changed and enhanced locally by adjusting the backgrounds, such as substrates, supporter, electrical gate, and so on. For the corrugated structure of monolayer graphene being placed on a periodically structured conducting substrate, the absorption can be pushed up to 100% in theory, which is determined by the frequencies of the plasmon resonances and the input light [150]. Graphene sheet covered on subwavelength dielectric gratings can also do this work [240]. A strong THz absorption peak was also observed in large scaled graphene due to the natural nanoscale inhomogeneities, such as substrate terraces and wrinkles [241]. More interestingly, inspired by the one-atom thick Luneburg lens and omnidirectional light absorber [49,242], a structure with the concept of "2D black hole" can be realized in graphene sheet, where the core of the graphene structure can capture and absorb the infrared light [243].

Although the enhancement by LSPR can be very efficient in THz and infrared freqiencies, light absorption in visible frequencies is still hard to be realized in graphene because of the properties of plasmons. Composite materials are adopted to solve this problem. For instance, the structure of graphene on one-dimensional photonic crystal has been proposed to improve the light absorption, where the enhancement indeed happen in the visible frequencies for both TM and TE modes of light [55]. However, the more general way is to hybridize graphene with metal plasmonic structures [244-247]. Plasmons in metals can resonate in visible frequencies, and in the hybrid structure the resonant frequency can be tuned by graphene to some degrees. From all of the above cases, graphene can be used in photoelectrics and photovoltaics due to the enhancement of light absorption.

*4.5. Surface-enhanced Raman scattering with graphene*

Raman spectroscopy is a powerful technique to provide detailed information of molecular structures and has proved to be a useful tool to analyze the structure of many materials, including graphene. There are many excellent reviews about Raman spectroscopy of graphene [248,249]. The obtained Raman spectra imply much intrinsic information about graphene [250,251]. Most importantly, the microanalysis of other molecule can be carried out with the enhancement effect from graphene substrate, which is named as graphene enhanced Raman scattering (GERS) [252-255]. The origin of GERS is mainly thought to be chemical mechanism, as plasmons in graphene are dominating in THz and infrared frequencies. However, the enhancement factor of GERS is not strong enough and hence limits graphene as a sensitive surface enhancement substrate. By combining graphene and metal nanoparticles, the SERS effect can be greatly improved, which is attribute to the electromagnetic interactions between metals and graphene. Therefore, composites of graphene with other metal plasmonic structure are often used for SERS studies. About 100 times of enhancement factor is observed for graphene covered by Ag nanoisland [209]; tens of times is found in graphene covered by Au-nanoparticle array [210] while it will increase to about 1000 for Au-nanoparticle array being located under graphene [256]. In addition, for the composite of graphene and Au film or nanoparticles, graphene can suppress the photoluminescence background so as to get clearer results [257]. Polarized plasmonic enhancement can also be realized in the composite graphene-metal particle structures [256,258]. From the achievements, the better strategy is setting graphene sheet on the top of metal plasmonic structures. In addition to the enhancement by chemical mechanism, electromagnetic field of the metal plasmonic structures can pass through the ultrathin graphene sheet and then enhance the Raman signals further, which make graphene/metal substrate to be a good choice for SERS. The flat surface and stability of graphene would also help for the uniform, stable, clean, and reproducible signals [254].

*4.6. Plasmonic detectors and sensors with graphene*

Graphene can be used as high performance detectors and sensors for room-temperature THz radiation [259], biomacromolecule [260,261], visible light [262], or even individual gas molecules [61]. With high efficient optical absorption in THz and infrared, graphene could be used as plasmonic detectors for light in those frequencies. The THz antennas based on graphene has already been discussed in section 4.1. With enhancement effect of SPs, the photodetectors and sensors hybridized graphene with other plasmonic structures are exploited for detecting light and optical sensing of macromolecules or biomolecules. Graphene in these devices is not the leading role, however, the advantages of tunability, ultrathin thickness, and biocompatibility still lead it to be a very important support in detecting and sensing.

Photovoltage and photodection based on graphene can also be enhanced by combining with plasmonic materials [245,247,262], with the enhancement of produced photocurrent. Fig. 26a shows that Au nanoparticles array is supported by back-gate graphene transistors, and the different IV characteristics can be obtained depending on the size of Au nanoparticles and the frequencies of the incident light [262]. Moreover, the enhancements of the photocurrent can reach up to 1500% as the thickness of the Au film increases to 12 nm. Generally, the enhancement peaks will redshift with increase of Au nanoparticles sizes. In addition to the monolayer, multicolor photodetection can also be realized in multilayer graphene based plasmonic structures. As shown in Fig. 26b, Au heptamer antenna structure is sandwiched between two monolayer graphene sheets which can be used as multicolor photodetector by adjusting the size of Au heptamer array [247]. The enhancement of the photocurrent can reach up to 800% for visible and near-infrared light, which was attributed to the transfer of hot electrons and SPs.

The quality of a sensor can be evaluated by four parameters: sensitivity, adsorption efficiency, selectivity and signal to noise ratio. The SPs resonant sensors exhibit promising potentials because of their very high sensitivity to the sensing medium, where the change in the refractive index of the sensed medium induces specific alternations in the characteristics of SPs resonances [261]. Generally, metal films are used as SPs resonant sensors, however, they are not the ideal choice because of the inactivation and the bad adsorption capacity. Metal film can be covered by graphene layers [263-265], where the adsorption for organic molecule or biomolecule can also be improved by $\pi-\pi$ stacking interaction because of the aromatic structure and the signal to noise ratio can be greatly enhanced. Moreover, the inactivation of SPs in metal can be eliminated effectively with the protection of graphene and the SPs will not be weakened due to the ultrathin thickness of graphene. A typical schematic of graphene based SPs resonant sensors is shown in Fig. 26c [266]. Au, Ag, Al *etc.* [267-269] can be used as plasmonic materials and the optimized film thickness (usually about 40-70 nm) is depending on the incident light and the background of the structure. According to theoretical calculation, Ag film is the best one due to the sharpest peaks of reflectance and sensitivity with respect to the incidence angle; Au film is the secondly best and then the others. The layer number of the graphene will also impact the sensitivity of the sensors, and fewer layers are beneficial to higher sensitivity. In order to get higher sensitivity, the structure can be further optimized. For instance, replacing the silica prism by chalcogenide prism [266], adding a silicon layer or a silica doped $B_2O_3$ between metal and graphene layers [63,270], and so on. However, the selectivity is the tender spot of the graphene based SPs resonant sensor, especially for the mixed sensing medium. Fixing acceptor on graphene for sensing specific molecular might be the sally port for better selectivity similar to aptasensors [62].

**Conclusions and outlooks**

SPs in Graphene has shown great advantages compared to conventional metal plasmonic materials, which including the extremely high confinement, relatively low loss in THz and infrared frequencies, high tunability, long relaxation time of electrons, and strong many-body interactions. Great achievements in graphene plasmonics have been made both on theoretical predictions and experimental applications. However, there are still some limitations and drawbacks for graphene used in plasmonics, especially in the propagation of plasmons. The lack of reliable THz light sources and THz detectors also limit the practical applications of graphene plasmonics, and smaller propagation length-to-wavelength ratio at high frequencies make graphene cannot outperform noble metals as a good platform for high frequencies SPPs [185]. Therefore, more efforts are expected on but not restricted to the following aspects:

1) The working frequencies of graphene SPs are in THz and infrared regions, and if such frequencies can be extended to near-infrared or even visible regions, it may find more potential applications. The possible ways are introducing extremely high and nondestructive doping in graphene, opening bandgaps, and making graphene-metal hybrid structures. Among them, the combination of graphene and conventional metal plasmonic materials seems more available. By taking the advantage of graphene's high tunability, such hybrid structure may present versatile properties in visible and near-infrared region and at same time with good tunability.

2) Although the propagation of SPPs in graphene has been demonstrated through nano-images by observing on the standing waves due to interference [52,53]. A direct experimental proof of the excitation, propagation, and detection of SPPs in graphene, similar to those in conventional metals, are highly demanded.

3) There are different kinds of plasmons in graphene, *i.e.*, 2D plasmons, $\pi$ and $\pi+\sigma$ plasmons [101]. The low energy 2D plasmons have been extensively studied, while those high energy plasmons are seldom addressed. Those high energy plasmons (*e.g.*, $\pi$ plasmons at 4-8 eV) may find potential applications in UV regions, *e.g.*, combing with materials that emit or absorb UV light.

4) The energy loss of SPs in graphene is expected to be further reduced, which would benefit its propagation and real applications.

5) The LSPR in graphene are indeed very attractive, which can affect the absorbance, transmission, and enhance the local electrical field, more experimental work are encouraged to be carried out in this subject.

6) The properties of graphene are not only affected by the structures and sizes, and different approaches have been demonstrated to tune the properties of graphene, including layer numbers, strain, defects, edges, stacking geometry [96,271-273], and so on. In turn, the properties of SPs in graphene can also be altered with the above approaches, but has not been paid attention to.

7) Another great advantage of graphene is its flexible nature [48], which would greatly enrich the graphene plasmonics. Flexible plasmonic devices based on graphene are possible due to the high confinement of SPs in graphene.

Although the abovementioned points are very challenge, once realized, they will certainly make graphene plasmonics one of the most fascinating fields. By designing the graphene based subwavelength plasmonic tunable devices, including THz plasmonic lasers, plasmonic antennas, plasmonic waveguides, Luneburg lens, modulators, polarizers and so on, the plasmonic circuit can be achieved which may launch a revolution in photonics and electronics.

**Acknowledgements**

This work was supported by the National Natural Science Foundation of China under Grant Nos. 51071045, 51271057, 61271057 and 11104026) the Program for New Century Excellent Talents in University of Ministry of Education of China under Grant No. NCET-11-0096 and NCET-11-0094, the Natural Science Foundation of Jiangsu Province, China, under Grant No BK2012757 and BK2011585, and the open

**Figures:**

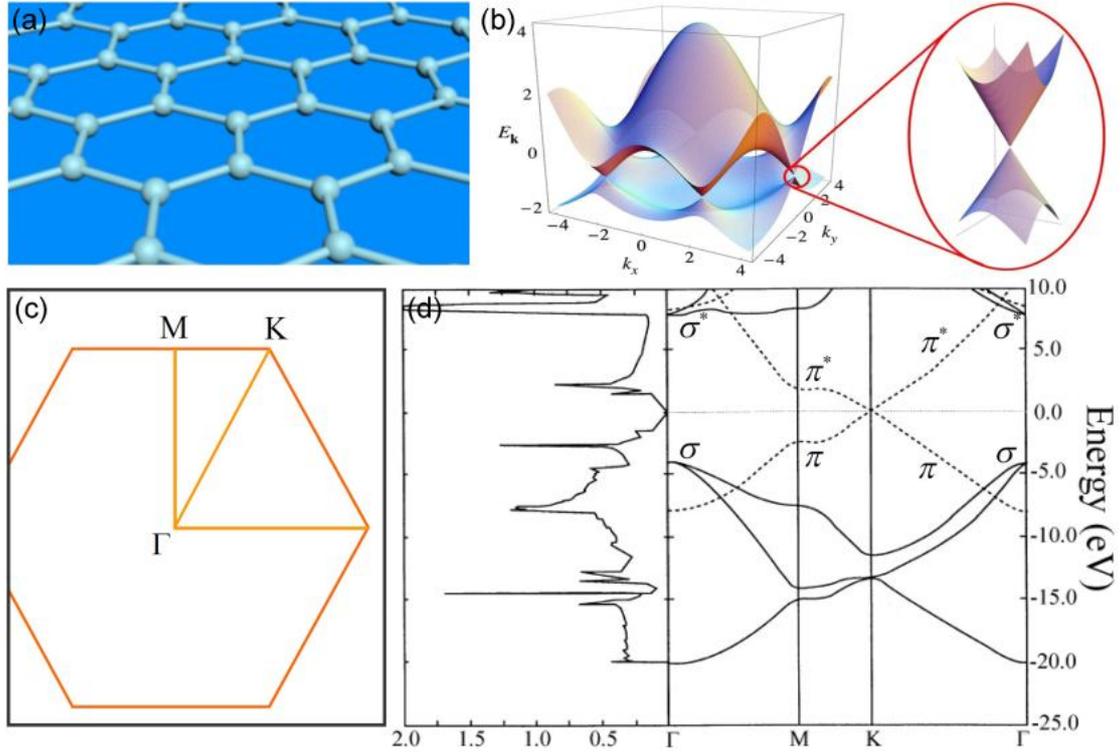

**Fig. 1.** (a) The molecule schematic of graphene. (b) The conical band structure with a zero-energy band gap at the Dirac point, where $t = 2.7\text{eV}$ and $t' = -0.2t$ [33]. (c) Schematic of graphene Brillouin zone. (d) Density of states (states/eV atom) and energy-momentum dispersion relation of $\pi$, $\pi^*$, $\sigma$ and $\sigma^*$ bands of monolayer graphene with the Fermi energy $E_F = 0$ [69].

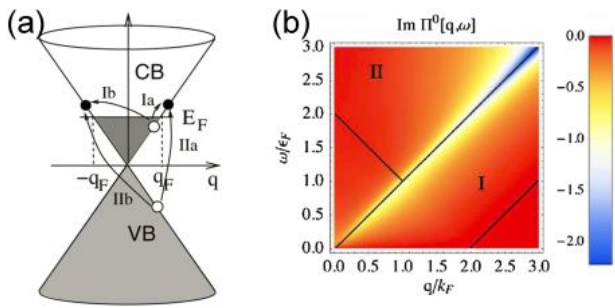

**Fig. 2.** Schematic of the SPE region for $n$-doped graphene. (a) Possible intraband transition (I) inside the conduction band and interband transition (II) between conduction band and the valence band respectively. (b) SPE regions and the spectral function corresponding to (a) [67].

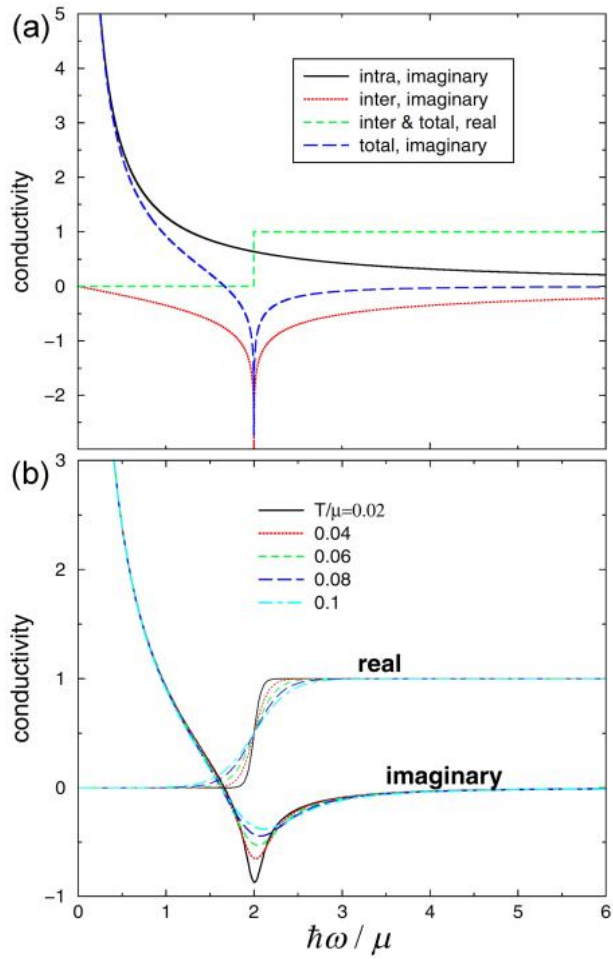

**Fig. 3.** The dynamic conductivity (unit $e^2/4\hbar$) of the graphene layer as a function of $\hbar\omega/\mu$ at zero temperature (a) and non-zero temperature (b) [89].

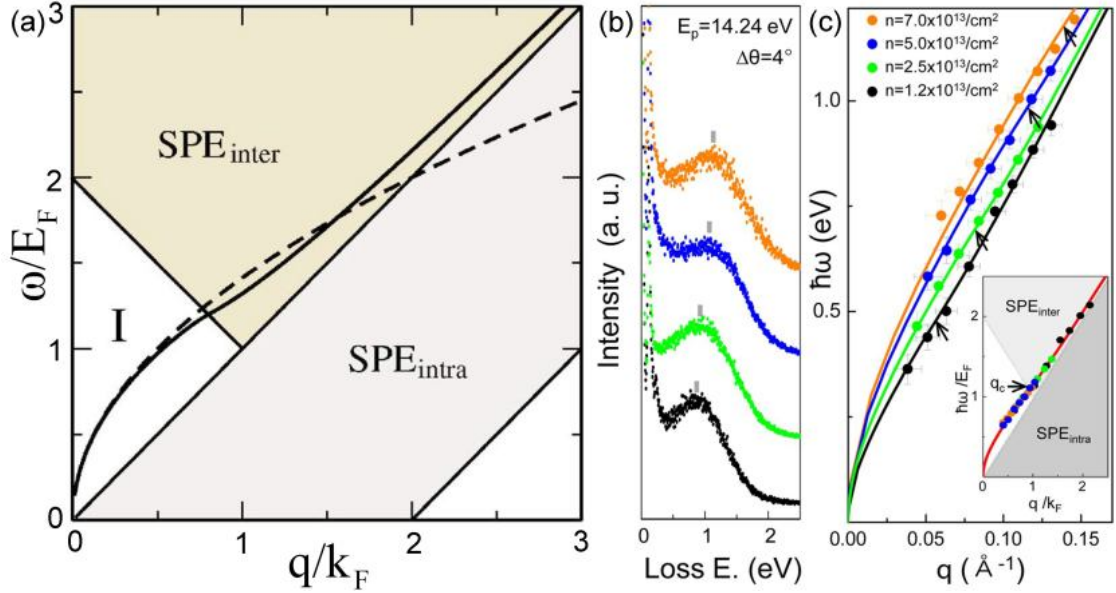

**Fig. 4.** (a) Dispersion relation of 2D electron gas (dashed line) and graphene (solid line) by RPA compared to the SPE region, where $\kappa = 2.5$, $\gamma = 6.5$ eVÅ$^{-1}$, and $n = 10^{12}$ cm$^{-2}$ [74]. (b) and (c) High-resolution EELS and dispersion relation of monolayer graphene epitaxially grown on SiC substrate at different carrier density from experiments [80].

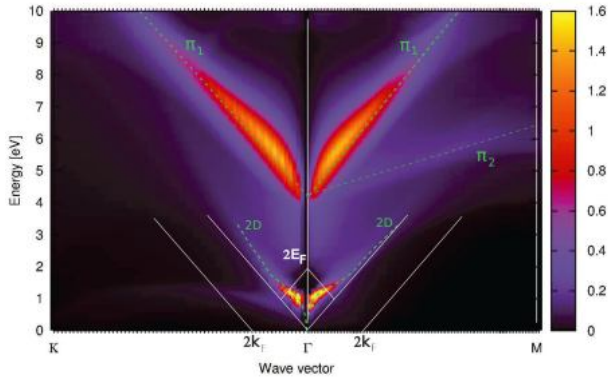

**Fig. 5.** Intensities of electronic excitations which also show the dispersion relations of 2D plasmons and $\pi$ plasmons, where $E_F = 1$eV, $\pi$ plasmons split into two kinds of modes in the $\Gamma - M$ direction [101].

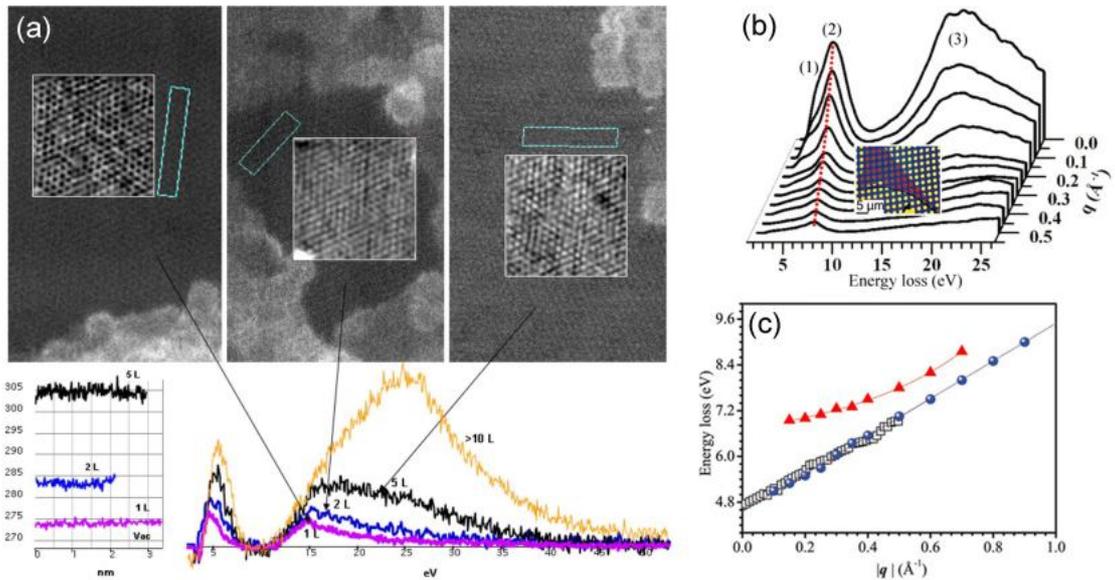

**Fig. 6.** EELS studies of free-standing graphene. (a) Three upper panels are high angle annular dark-field scanning transmission electron microscope images of one, two and five layers of graphene; the two lower images show the intensity (left panel) along the long dimensions of the rectangular cyan-framed boxes and EELS spectra (right panel) of one, two, five and several layers of graphene showing $\pi$ and $\pi+\sigma$ plasmon in the three upper panels [102]. (b) Background-subtracted EELS spectrum of monolayer graphene. The features labeled (1), (2) and (3) are attributed to $\pi \rightarrow \pi^*$ interband transitions, $\pi$ and $\pi+\sigma$ plasmon excitations, respectively. Inset: A false colour TEM bright field image of the graphene layers (red) on the Carbon grid with the holes colered in yellow [109]. (c) Experimental dispersion of the $\pi$ plasmon peak of (b) (open squares), compared with vertically aligned SWCNT (filled blue spheres) and graphite (filled red triangles). And the fits of the dispersion curves (dotted curves) [109].

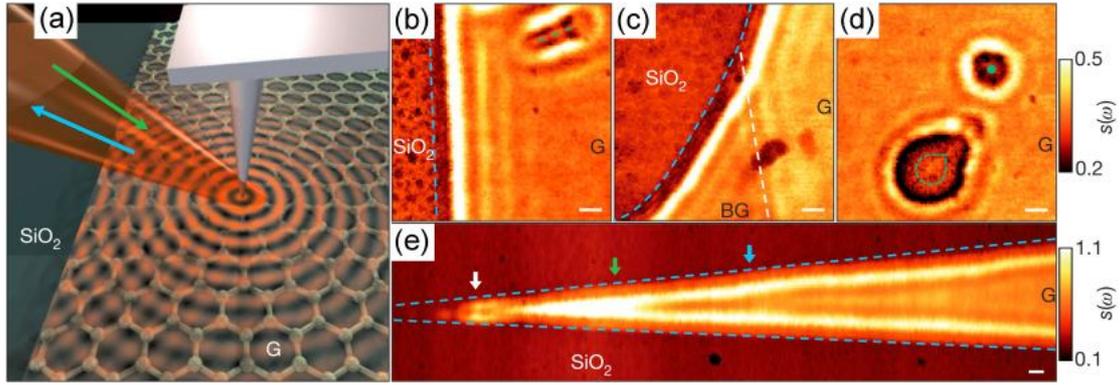

**Fig. 7.** Infrared light nano-imaging of SPs in graphene based on $SiO_2$. (a) Diagram of an infrared nano-imaging of SPs with the green and blue arrows indicating the incident and back-scattered light respectively. Concentric red circles illustrate plasmon waves induced by the illuminated tip of AFM. (b-e) Images of infrared amplitude and the interference pattern close to graphene edges (blue dashed lines) and defects (green dashed lines and green dot), and at the boundary between single (G) and bilayer (BG) graphene (white dashed line). Scale bars, 100 nm. [52].

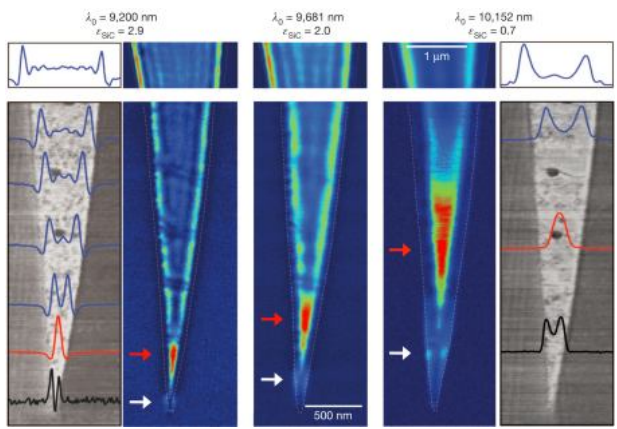

**Fig. 8.** Optical nano-imaging of SPs in tapered graphene ribbon on the carbon-terminated surface of 6H-SiC. Fringes SPs formed by standing wave can be tuned by the incident light and the dielectric constant of substrate [53].

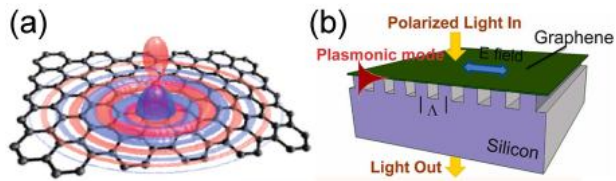

**Fig. 9.** (a) Near electric field produced by a perpendicular dipole. The red (blue) regions in the three-dimensional contour indicate the real (imaginary) part of the perpendicular electric field [39]. (b) Schematics of SPs excited by a silicon diffractive grating [127].

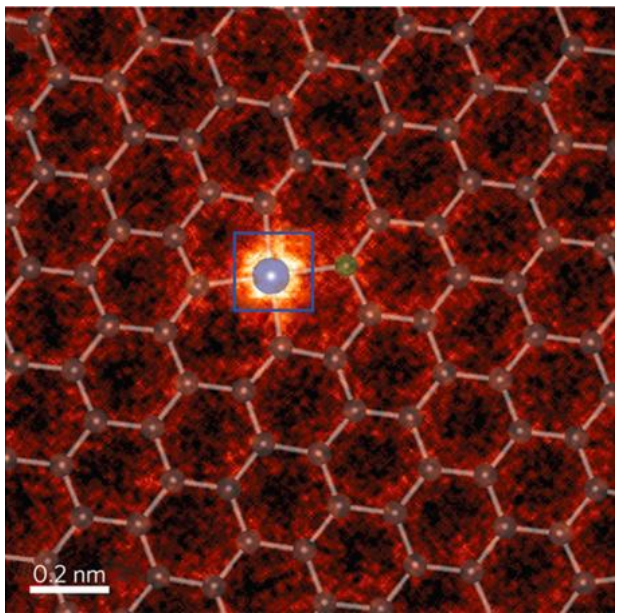

**Fig. 10.** Annular dark-field survey image, EELS imaging and overlaid structural model with atoms of carbon (gray), silicon (blue) and nitrogen (green), where electrons lead to LSPR at atomic scale [132].

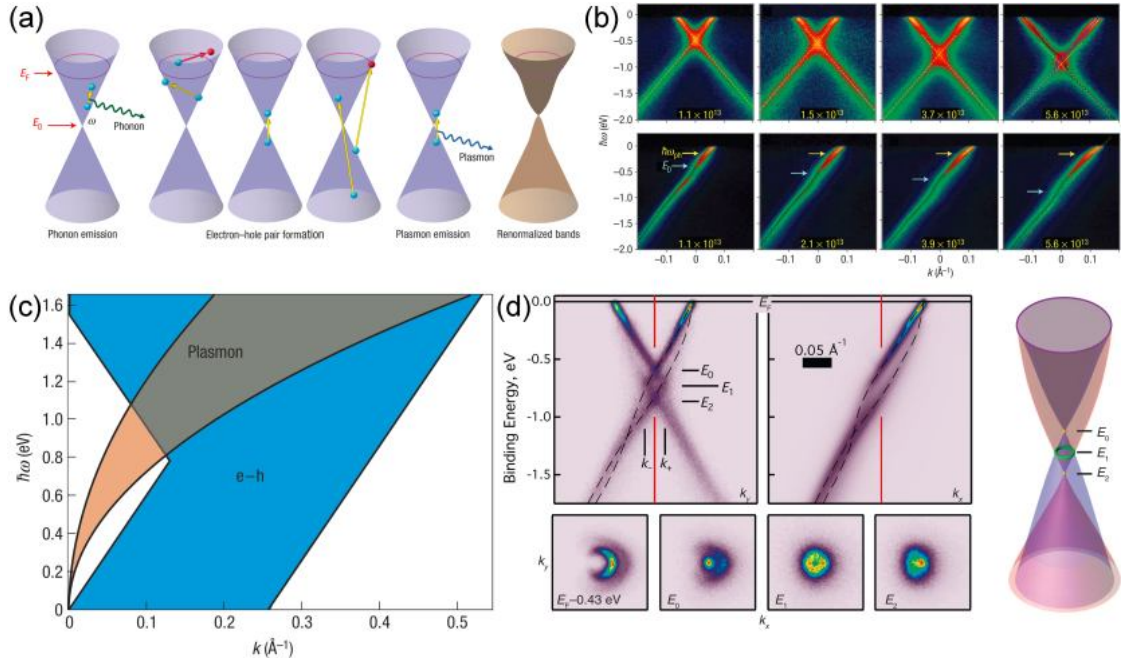

**Fig. 11.** (a) Schematic of electron-phonon, electron-electron and electron-plasmon interactions in doped graphene, where only the magenta arrows indicate the successful transitions, the rightmost panel shows the renormalized bands distorted by these processes [135]. (b) Experimental energy band by ARPES around the Dirac point of doped graphene based on SiC substrate, with different electron concentration (cm$^{-2}$), The kinked dispersion of the bands together with linewidth variations are clearly visible in the fitted peak positions (dotted lines) [135]. (c) plasmon dispersion (pink) and electron-hole pair excitations region for graphene with electron concentration $5.6 \times 10^{13}$ cm$^{-2}$, where plasmon dispersion calculated for a range of dielectric constants from 3 (upper edge) to 10 (lower edge) [135]. (d) ARPES study of plasmaron in doped graphene with electron concentration $1.7 \times 10^{13}$ cm$^{-2}$, where the Dirac point is resolved into two point and one ring, dashed guide lines indicate the plasmaron band (lower pass through $E_2$ point) and the pure charge band (upper) [50].

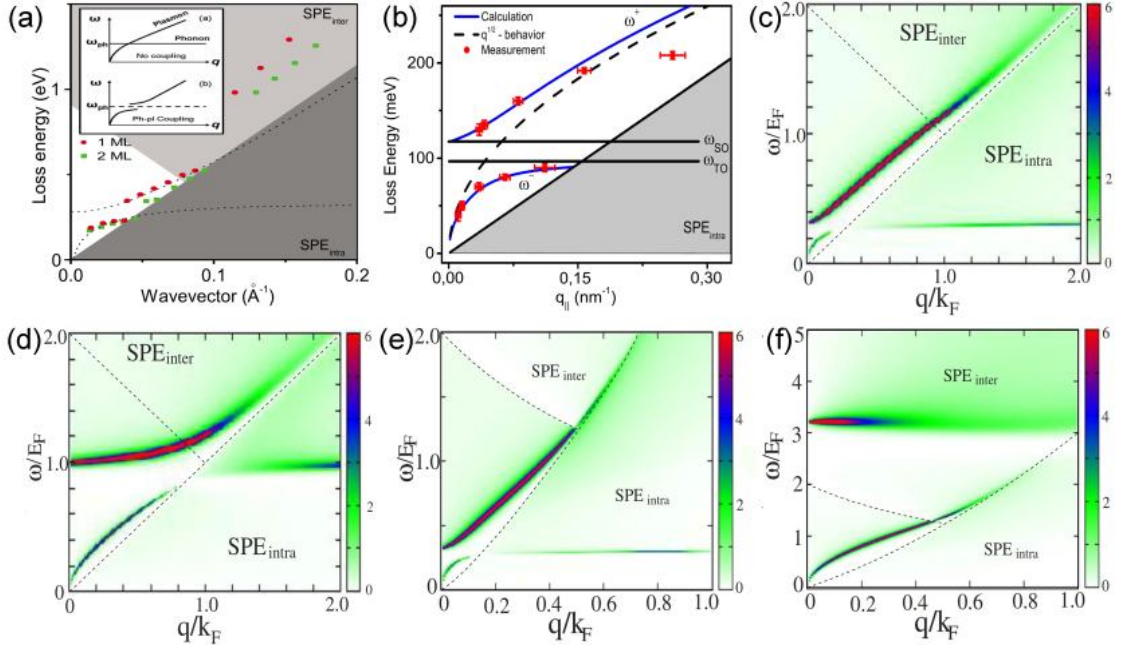

**Fig. 12.** Dispersion of the plasmon-phonon coupled modes. (a) The energy loss peaks from angle-resolved EELS experiments for single-layer (1 ML) and bilayer (2 ML) graphene. Inset: schematic of the coupling [38]. (b) Dispersion of the coupled plasmon-phonon modes with a gap between $\omega^{\pm}$ modes. The red dots with error bars are experimental data [37]. (c-f) The strength of plasmon-phonon coupled modes in single-layer graphene based on polar substrate for electron concentration of $10^{13}$ cm$^{-2}$ (c) and $10^{12}$ cm$^{-2}$ (d), and in bilayer graphene for electron concentration of $10^{13}$ cm$^{-2}$ (e) and $10^{12}$ cm$^{-2}$ (f) respectively [143].

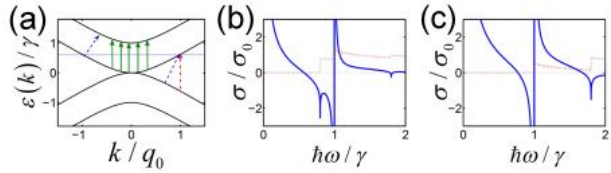

**Fig. 13.** (a) Schematic of the band-structure of bilayer graphene and some possible interband transtions (arrows), where $q_0 = \gamma / \hbar v_F$. The real (red dotted lines) and imaginary (blue solid lines) part of the conductivity of graphene for different doping of $\mu = 0.4\gamma$ (a), and $\mu = 0.9\gamma$ (b), where $\sigma_0 = e^2 / 2\hbar$ [153].

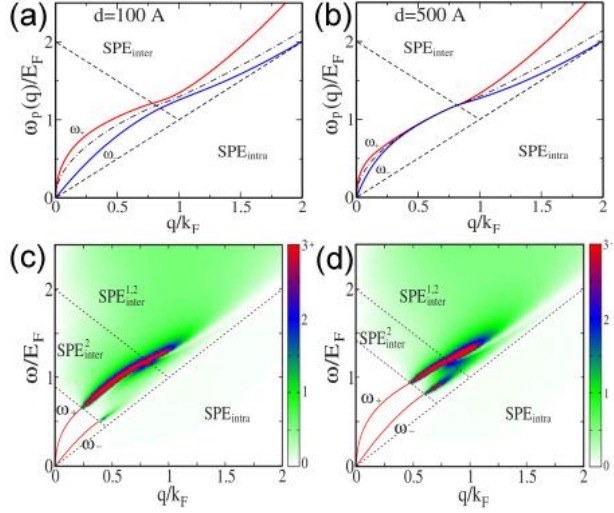

**Fig. 14.** Dispersion of the bilayer graphene at different separated distance $d$ and electron concentrations. In (a) and (b), $n_1 = n_2 = 10^{12}$ cm$^{-2}$ and the dot-dashed line indicates the plasmon mode dispersion of SLG with the same density. (c) and (d) show the strength of energy loss related to the SPE regions where $d = 100$ Å and $n_1 = 10^{12}$ cm$^{-2}$, $n_2 = 0.2 n_1$ for (c) and $n_2 = 0.5 n_1$ for (d) [156].

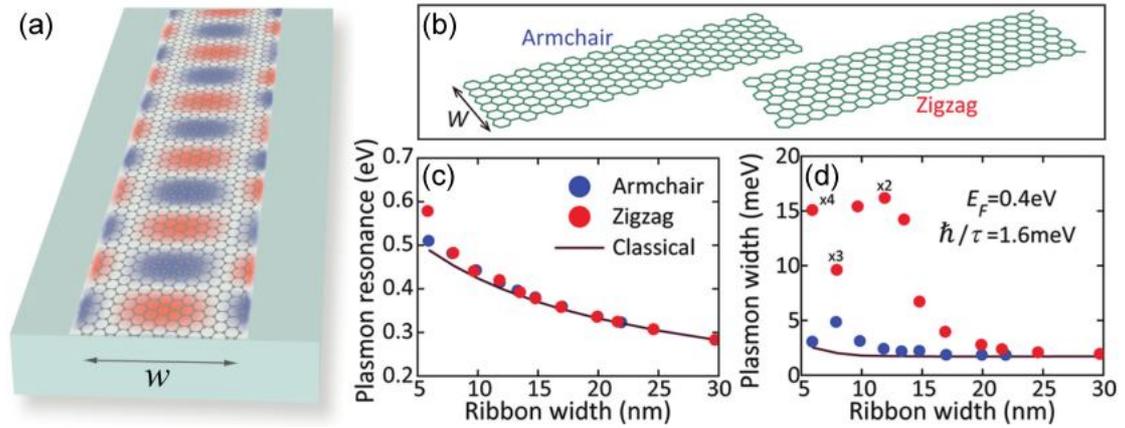

**Fig. 15.** (a) Schematic of SPs propagate along GMR, the color plot presents an example of the electric field and show the waveguiding and edge modes, $w$ is the width of the graphene ribbon [165]. (b) Schematic of zigzag and armchair GNRs. (c) The relationship between plasmon energy and ribbon width from RPA (symbols) and classical theory (solid line). (d) The relationship between plasmon half-area width and ribbon width [151].

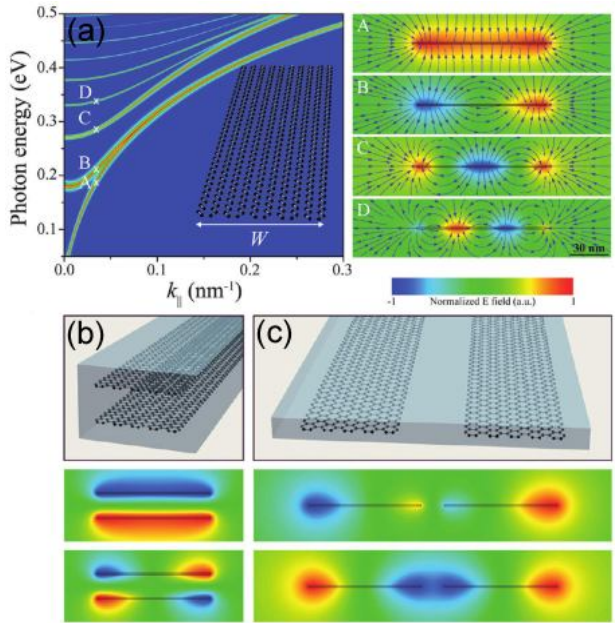

**Fig. 16.** (a) Dispersion (left panel) of a GNR of $w = 100$ nm and $E_F = 0.5$ eV and the real part of the electric field amplitude (right panels) along the ribbon direction corresponding to modes labeled A-D in the dispersion diagram for $k_\parallel = 0.035$ nm$^{-1}$. (b) and (c) Two different hybridization (upper panel) with two GNRs and the symmetric and antisymmetric coupling of plasmons (lower panels) in the ribbons [148].

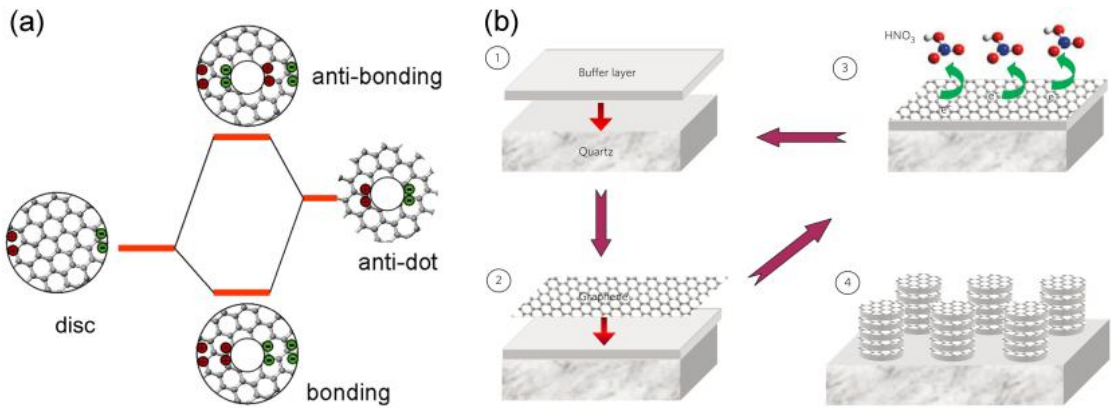

**Fig. 17.** (a) Energy level diagram for the plasmon hybridization of a disk and an antidot [126]. (b) Three layer deposition steps and a single lithographic step are used to produce stacked structures: 1) polymer buffer layer coating; 2) graphene deposition; 3) graphene doping and 4) lithographic patterning (e.g. patterned disk arrays). Repeating step 1)-3) makes any layers of graphene [42].

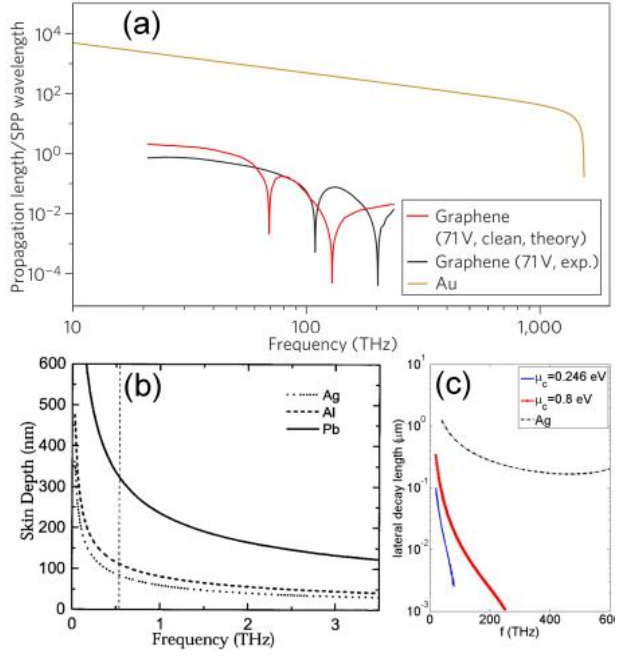

**Fig. 18.** (a) $\delta_{SPP}/\lambda_{SPP}$ with respect to frequency of incident light of graphene and Au [185]. The results for Au are for a 30-nm-thick film at room temperature. And the results for graphene are for the highly electrical doped case calculated from the experimental data in reference [186] and from the theoretical data in reference [187]. (b) Frequency-dependent penetration depths of SPPs in Pb, Al and Ag based on Si substrate, where $\varepsilon_d = 11.68$ for silicon [188]. (c) The lateral decay length (i.e., vertical decay length) of SPPs in graphene ($\mu_c$ indicates the chemical potential here) and in 30 nm thick Ag film [48].

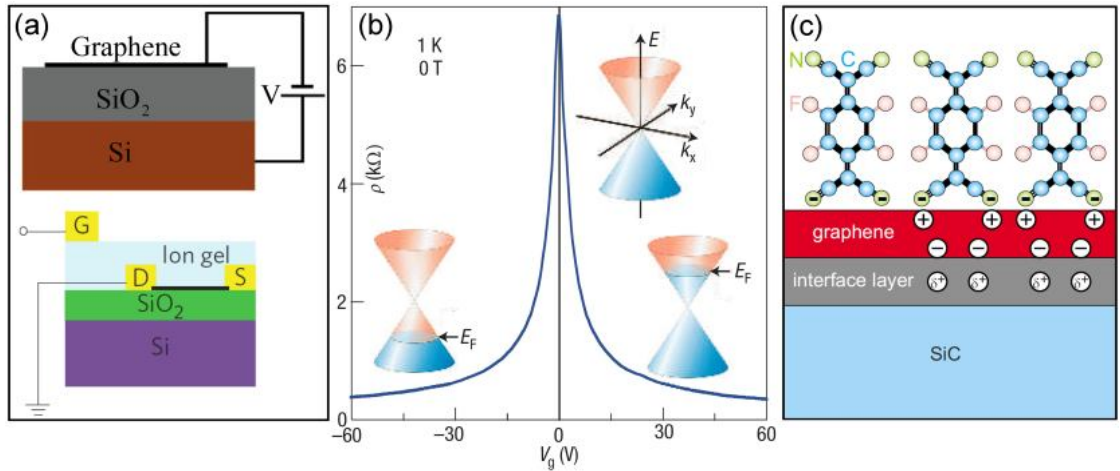

**Fig. 19.** (a) Electrical doping: bias voltage applying (upper panel) and ion-gel top gate (lower panel) [41]. (b) Doping in monolayer graphene, p- and n-doping according to the left and right parts around $V_g = 0$, respectively [66]. (c) Chemical doping by the electron acceptor of F4-TCNQ [160].

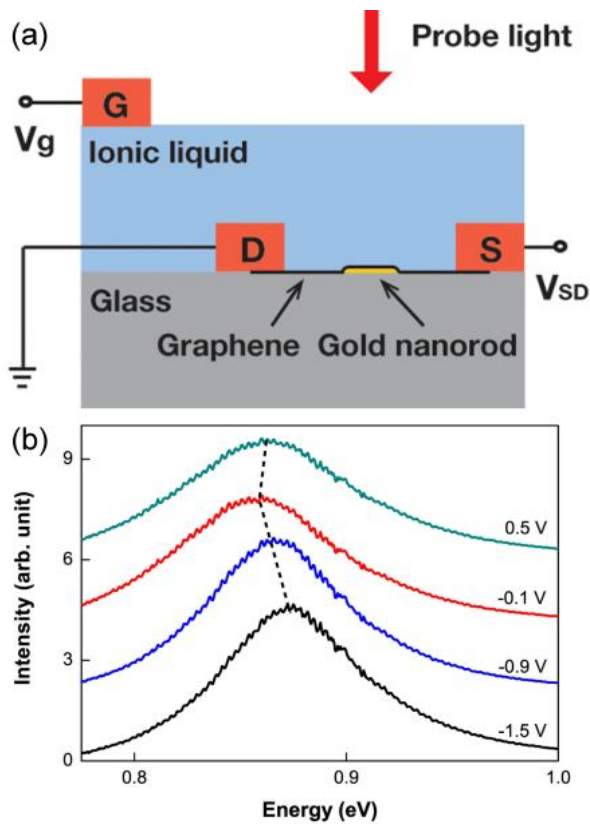

**Fig. 20.** (a) Schematic of a typical device of gold nanorods covered by graphene, in which the SPs resonances are controlled by a top electrolyte gate with ionic liquid. (b) Reyleigh scattering spectra of the device at different gate voltage, which clearly demonstrate the capability to control SPs [207].

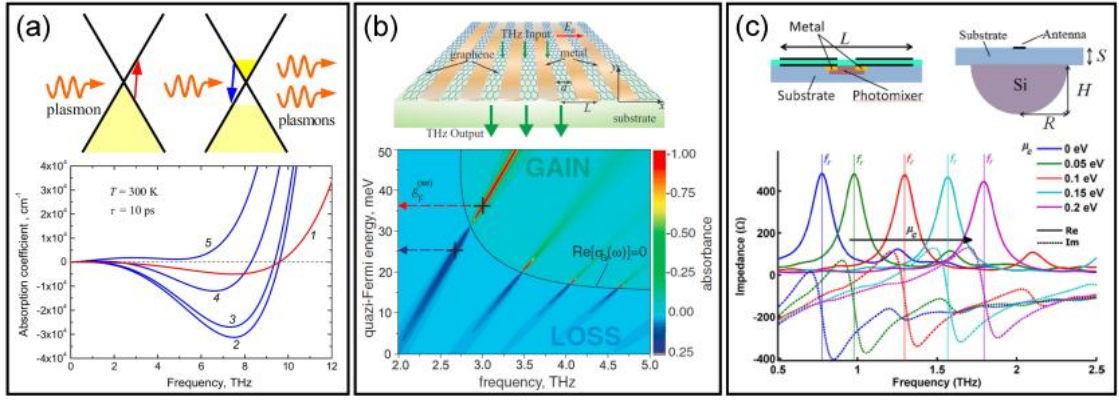

**Fig. 21.** (a) Graphene THz plasmonic amplifier: the upper panel showing interband population inversion by the stimulated absorption of plasmons and stimulated emission of plasmons; the lower panel showing the surface plasmon absorption for monolayer graphene based on different substrate with the refraction indices of $n_1 = 1.0$, $n_2 = 3.4$, $n_3 = 3.4 + 0.01i$, $n_4 = 3.4 + 0.05i$ and $n_5 = 3.4 + 0.1i$ respectively, where $E_F = 20$ meV and the negative absorbance indicates plasmon gain [218]. (b) A planar array of graphene resonant micro/nanocavities: the upper panel showing the schematic view; the lower panel showing the THz wave absorbance situation at the given parameters of $L = 4\,\mu m$, $a = 2\,\mu m$ and $\tau = 1\,ps$, where the blue and red arrows mark the Fermi energies for the maximal absorption and plasmonic lasing at the fundamental plasmon resonance respectively [217]. (c) Graphene dipole plasmonic antenna: the upper panels showing cross view of the plasmonic graphene dipole antenna (left), and structure including the silicon lens for better directivity (right); the lower panel showing the real and imaginary parts of the input impedance tuned by chemical potential at THz frequencies. Each dipole arm is a set of two stacked graphene patches separated by a thin $Al_2O_3$ ($\varepsilon_r = 9$), the antenna width is 7 μm and the total length $L = 11\,\mu m$, the two dipole arms lies on a GaAs dielectric substrate ($\varepsilon_r = 12.9$) with a 2 μm THz photomixer between them, other parameters of the structure are $S = 160\,\mu m$, $H = 572\,\mu m$, $R = 547\,\mu m$, and $\varepsilon_r = 11.66$ respectively [220].

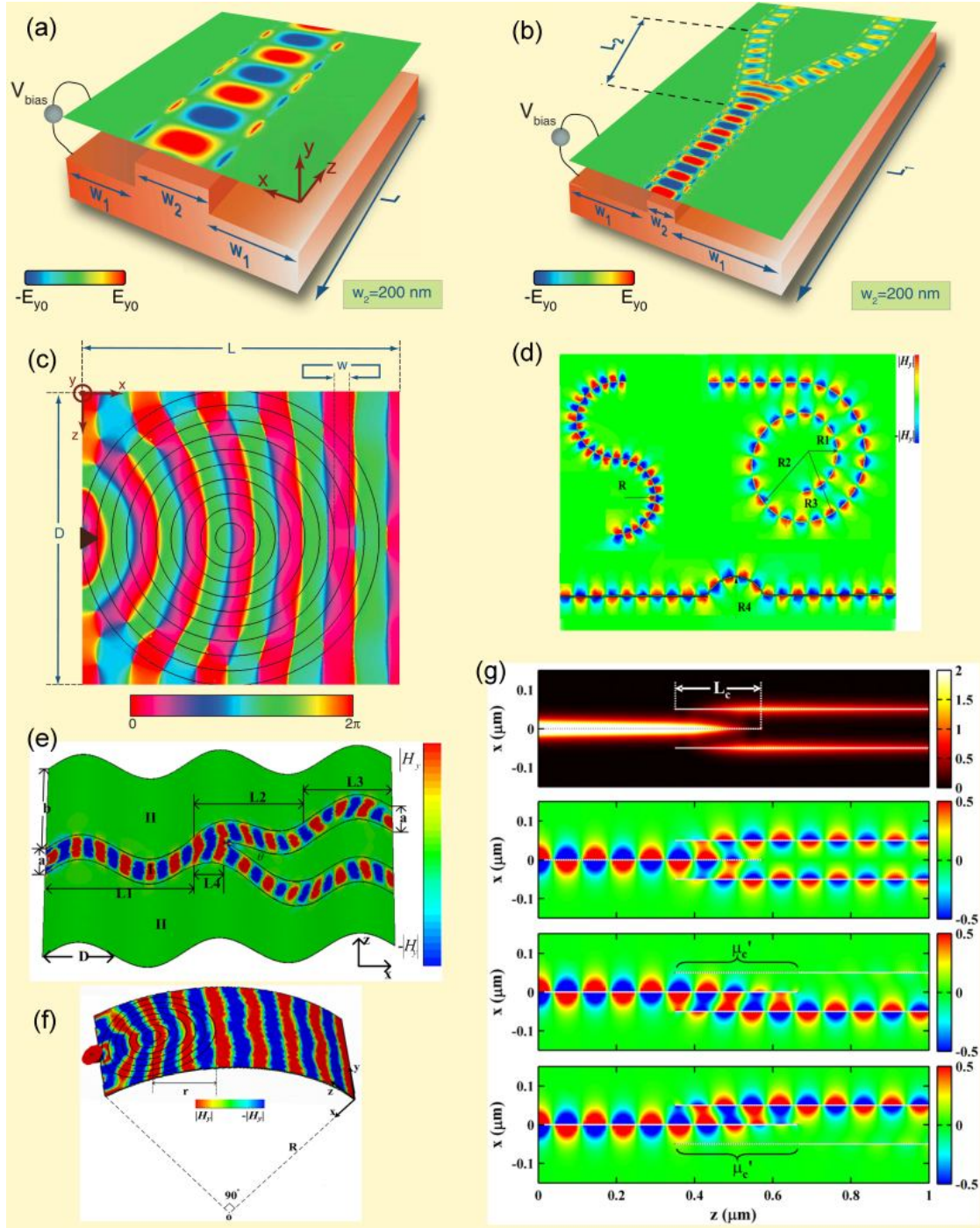

**Fig. 22.** (a) Simulation of electric field of TM SPPs at $f = 30$ THz in $y$ direction in graphene sheet based on an uneven ground plane, between which is the dielectric spacer ($\mu = 0.15$ eV, $L = 560$ nm, $w_1 = 200$ nm). (b) Similar to (a), but the waves split into two paths ($L_1 = 1077$ nm, $L_2 = 560$ nm, $w_1 = 600$ nm). (c) Similar to (a), but the ring shape graphene based on even ground plane which performs as a one-atom thick Luneburg lens, ($D = 1.5$ μm, $w = 75$ nm, $L = 1.6$ μm) [43]. (d), (e) and (f) show the simulation results of the tangent magnetic fields for SPP waves in the curved flexible graphene structures [48]. (g) Intensity distribution of SPPs in graphene waveguide splitter (upper panel), and electric field perpendicular to the graphene sheet when no bias, bias applied on the upper arm, and bias applied on the lower arm respectively (the lower three panels), where $L_c = 221.4$ nm, $\mu'_c = 0.05$ eV, and $\hbar\omega = 0.124$ eV [163].

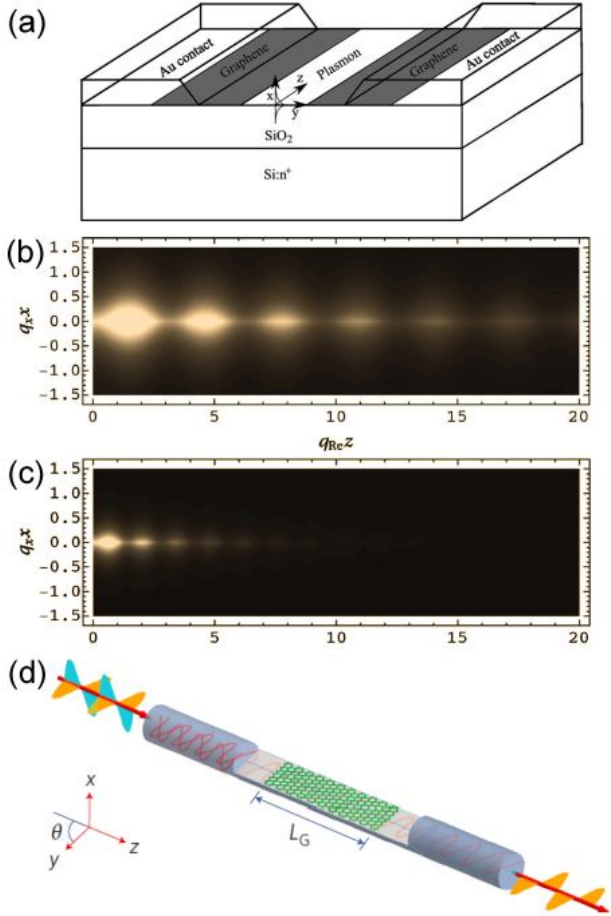

Fig. 23. (a) A plasmonic modulator with a gate bias being applied to the graphene monolayer through Au electrode and Si: $n^+$ substrate, where SPs propagate along the sheet. (b) and (c) are simulated results of (a), propagation of $\lambda_0 = 10\,\mu$m TM mode SPs in the modulator tuned by the electrical gating, the carrier concentration of the upper/low panel is $2.41 \times 10^{12} / 3.8 \times 10^{11}$ cm$^{-2}$ [232]. (e) Schematic of an in-line fiber-to-graphene polarizer based on a side-polished optical fiber, where $L_G$ is the propagation length and the polarization angle $\theta$ is defined as the angle between the polarization direction and the graphene plane [60].

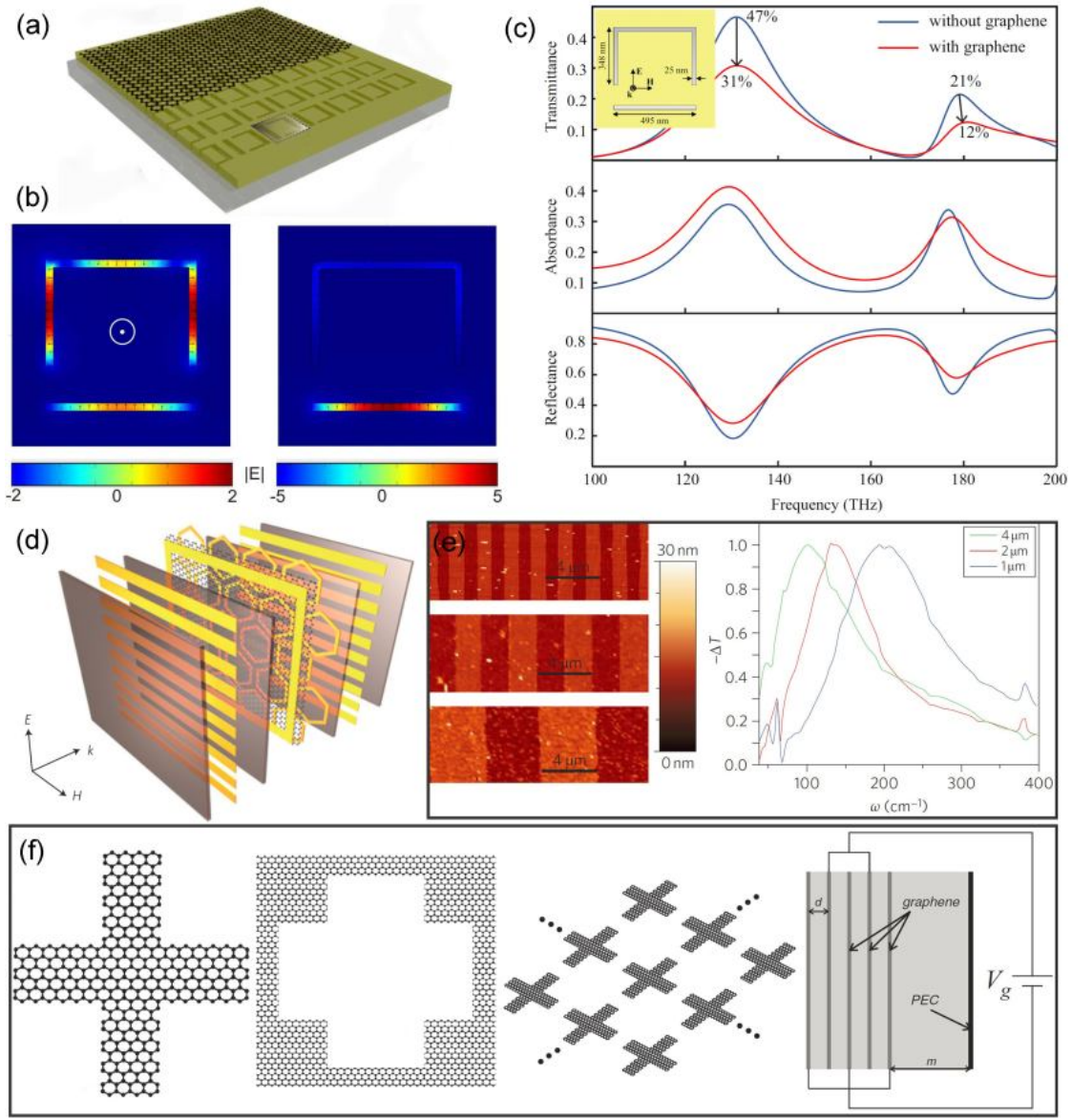

**Fig. 24.** (a) Asymmetrically split ring made by Au based on $Si_3N_4$ membrane whose top is covered by a monolayer graphene [45]. (b) Simulated electric field maps of (a) at the trapped-mode (left) and the dipole resonance (right) [45]. (c) Simulated results of transmittance, absorbance, and reflectance of a Fano ring covered with and without graphene [236]. (d) Schematic of a gate-controlled active graphene metamaterial composed of a monolayer graphene on a layer of hexagonal metallic meta-atoms contacted with extraordinary optical transmission electrodes [238]. (e) Plasmonic THz metamaterials made by graphene micro-ribblons, the structure is shown by the AFM imaging (left), the absorption peak is observed relative absorbance spectrum which is impacted by the structure size [41]. (f) Periodic structured shape and antistructured

shape graphene can be stacked as the metamaterial and can be controlled by bias voltage, cross shape, anticross shape, and the cross shaped array are shown [234].

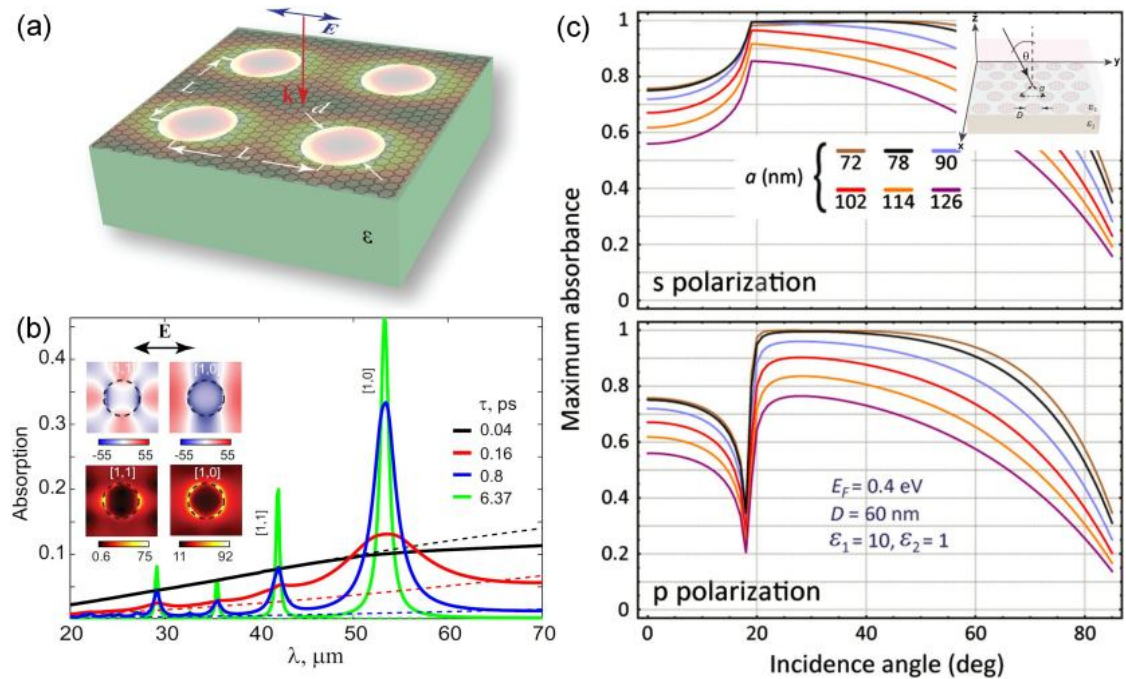

**Fig. 25.** (a) and (b) Enhanced absorption in the graphene sheet with periodic antidots array, the dashed lines indicate the spectra for graphene monolayer, where $\varepsilon = 1$, $\mu = 0.2$ eV, $L = 2d = 5$ μm; the insets of (b) show the modulus of the spatial distribution of the electric field (below) and the real part on the direction of the incident wave electric field shown by an arrow (above) respectively [174]. (c) A completely optical absorber made by the graphene with periodic nano-disks array. For continuous frequencies input light, there exit a absorption peak for the fixed structure and input angle, and the lower two panels show the absorption peaks for the different structures and different input angles of *s*-polarized and *p*-polarized light [56].

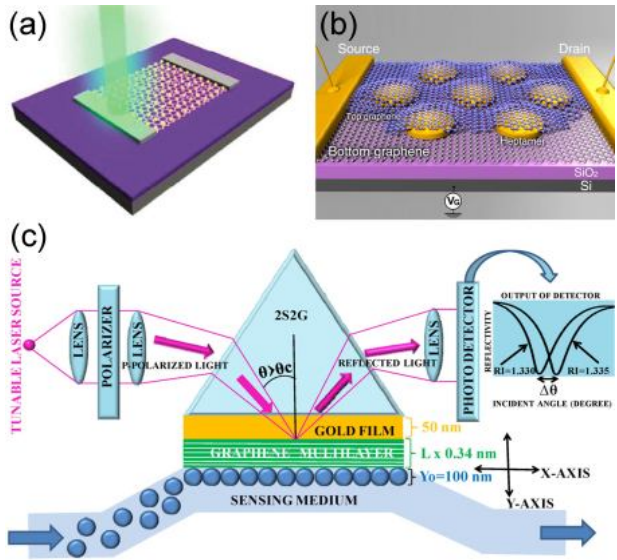

**Fig. 26.** (a) Schematic of graphene photodetector, with a laser scanning across the graphene-metal junction [262]. (b) Schematic of one single gold heptamer sandwiched between two monolayer graphene sheets, where the bias is used to electrically dope the graphene [247]. (c) Schematic of experimental setup for plasmonic biosensor with chalcogenide prism, gold and graphene multilayers [266].

# Tables

## Table 1
TM and TE surface plasmon in graphene.

| Mode | $\sigma''$ | Contribution | Frequency region | Damping |
|---|---|---|---|---|
| **TM** | > 0 | Intraband | THZ, far-infrared | No |
| **TE** | < 0 | Interband | Far- and near-infrared | Finite |

## Table 2
Energy peaks of four Carbon materials related to $\pi/\pi+\sigma$ plasmon and the interband transition from spectroscopy experiments or theoretical predictions..

| | Plasmon (eV) | | Transition (eV) | | | |
|---|---|---|---|---|---|---|
| | $\pi$ | $\pi+\sigma$ | $\pi \to \pi^*$ | $\sigma \to \pi^*$ | $\pi \to \sigma^*$ | $\sigma \to \sigma^*$ |
| **Graphite**[102,103] | 7 | 26 | 5-7 | 10-15 | 10-15 | 15-20 |
| **Fullerene**[104-106] | 6 | 25 | 2-5 | 13.5 | | 14.5 |
| **SWCNT**[103,107,108] | 4.5 | 15 | 4.2 | 11-12 | 14-15 | 11.4 |
| **Graphene**[102,109] | 4.7 | 14.6 | 4 | | | >10 |

## Table 3
Effective screening constants for graphene at different substrates [138].

| Substrate | $\Delta E$ | $\delta k$ | $\alpha_G$ | $\varepsilon$ | $\varepsilon_r$ |
|---|---|---|---|---|---|
| **Au-SiC** | $0.12 \pm 0.04$ | $0.09 \pm 0.02$ | $0.05 \pm 0.01$ | $44 \pm 9$ | $87 \pm 18$ |
| $6\sqrt{3}$ **C-SiC** | $0.21 \pm 0.02$ | $0.16 \pm 0.01$ | $0.1 \pm 0.04$ | $22 \pm 8$ | $43 \pm 16$ |
| **F-SiC** | $0.40 \pm 0.05$ | $0.29 \pm 0.03$ | $0.4 \pm 0.05$ | $5.5 \pm 0.7$ | $10 \pm 1.3$ |
| **H-SiC** | $0.49 \pm 0.02$ | $0.34 \pm 0.01$ | $0.5 \pm 0.03$ | $4.4 \pm 0.3$ | $7.8 \pm 0.5$ |

**Table 4**

The expressions of key parameters of SPP for metals and highly doped graphene [18,19,39,182], where $\alpha \approx 1/137$ is the fine structure constant and $\varepsilon_r$ is the dielectric constant of the substrate.

| Parameter | Metal | Graphene |
|---|---|---|
| **Wavelength ($\lambda_{SPP}$)** | $\lambda_0 \left[ (\varepsilon_d + \varepsilon'_m)/\varepsilon_d \varepsilon'_m \right]^{1/2}$ | $\lambda_0 (4\alpha E_F)/\hbar\omega(\varepsilon_r + 1)$ |
| **Propagation distance ($\delta_{SPP}$)** | $\lambda_0 (\varepsilon'_m)^2 \left[ (\varepsilon_d + \varepsilon'_m)/\varepsilon_d \varepsilon'_m \right]^{3/2} / 2\pi\varepsilon''_m$ | $\lambda_0 (\tau\alpha E_F)/\pi\hbar(\varepsilon_r + 1)$ |
| **Penetration depth ($\delta_i$)** | $\delta_d = \lambda_0 \left| (\varepsilon_d + \varepsilon'_m)/\varepsilon_d^2 \right|^{1/2}/2\pi$ <br> $\delta_m = \lambda_0 \left| (\varepsilon_d + \varepsilon'_m)/\varepsilon'^2_m \right|^{1/2}/2\pi$ | $\lambda_{SPP}/2\pi$ |

**Table 5**

The relative dielectric constants of Au, Ag, Cu and Al at different optical frequencies calculated from the experimental data in reference [184].

| $\lambda$/nm | Au | Ag | Cu | Al |
|---|---|---|---|---|
| **413.3** | -1.16+6.41$i$ | -4.22+0.73$i$ | -3.49+5.22$i$ | -24.93+5.25$i$ |
| **652.6** | -9.89+1.05$i$ | -17.20+1.16$i$ | -13.42+1.57$i$ | -58.93+23.30$i$ |
| **826.6** | -29.02+2.03$i$ | -30.23+1.60$i$ | -27.60+2.74$i$ | -61.55+45.54$i$ |
| **1240** | -76.77+6.52$i$ | -71.97+5.59$i$ | -71.38+7.33$i$ | -154.8+30.25$i$ |
| **2480** | -238.8+37.36$i$ | -227.1+28.36$i$ | -172.9+30.36$i$ | -645.9+157.2$i$ |